\documentclass[letterpaper,twocolumn,showpacs,amsmath,amssymb,showkeys]{revtex4}

\usepackage{graphicx,color}
\usepackage[colorlinks,plainpages=false,linkcolor=black,urlcolor=blue,citecolor=blue,pdfpagemode=UseNone,pdfstartview=FitH]{hyperref}

\begin{document}

\title{Iron based superconductors: magnetism, superconductivity and electronic structure\\(Review Article)}

\author{A.~A.~Kordyuk}
\affiliation{Institute of Metal Physics of National Academy of Sciences of Ukraine, 03142 Kyiv, Ukraine}

\begin{abstract}
Angle resolved photoemission spectroscopy (ARPES) reveals the features of the electronic structure of quasi-two-dimensional crystals, which are crucial for the formation of spin and charge ordering and determine the mechanisms of electron-electron interaction, including the superconducting pairing. The newly discovered iron based superconductors (FeSC) promise interesting physics that stems, on one hand, from a coexistence of superconductivity and magnetism and, on the other hand, from complex multi-band electronic structure. In this review I want to give a simple introduction to the FeSC physics, and to advocate an opinion that all the complexity of FeSC properties is encapsulated in their electronic structure. For many compounds, this structure was determined in numerous ARPES experiments and agrees reasonably well with the results of band structure calculations. Nevertheless, the existing small differences may help to understand the mechanisms of the magnetic ordering and superconducting pairing in FeSC.
\end{abstract}


\maketitle

\tableofcontents

\section{Introduction}

Four years ago, the discovery of LaO$_{1-x}$F$_x$FeAs \cite{KamiharaJACS2008}, a new superconductor with transition temperature at 26 K, has marked the beginning of a new era in superconducting research. The Copper Age has been replaced by the Iron Age, i.e. all the researchers and fundings have switched from the high-$T_c$ cuprates (HTSC or CuSC) to the iron based superconductors (FeSC), as it is clear from a number of early reviews on the subject \cite{SadovskiiPU2008, IvanovskiiPU2008, IzyumovPU2008, IshidaJPSJ2009, MazinPhC2009, JohnstonAPh2010, PaglioneNPh2010}. Today, after four years of active research, Ref.\,\onlinecite{KamiharaJACS2008} has been cited more than 3000 times and the investigation of FeSC is in the mainstream of the condensed matter physics \cite{WenARoCMP2011, StewartRMP2011, HirschfeldRPP2011, ChubukovAR2012}.

There are several good reasons why the FeSC are so interesting. First, they promise interesting physics that stems from a coexistence of superconductivity and magnetism. Second, providing much larger variety of compounds for research and having multi-band electronic structure, they give hopes to resolve finally the mechanism of high temperature superconductivity and find the way of increasing $T_c$. Lastly, the FeSC are quite promising for applications. Having much higher $H_c$ than cuprates and high isotropic critical currents \cite{PuttiSST2010, MollNM2010, GurevichRPP2011}, they are attractive for electrical power and magnet applications, while the coexistence of magnetism and superconductivity makes them interesting for spintronics \cite{PatelAPL2009}.

\begin{figure}
\begin{center}
\includegraphics[width=0.45\textwidth]{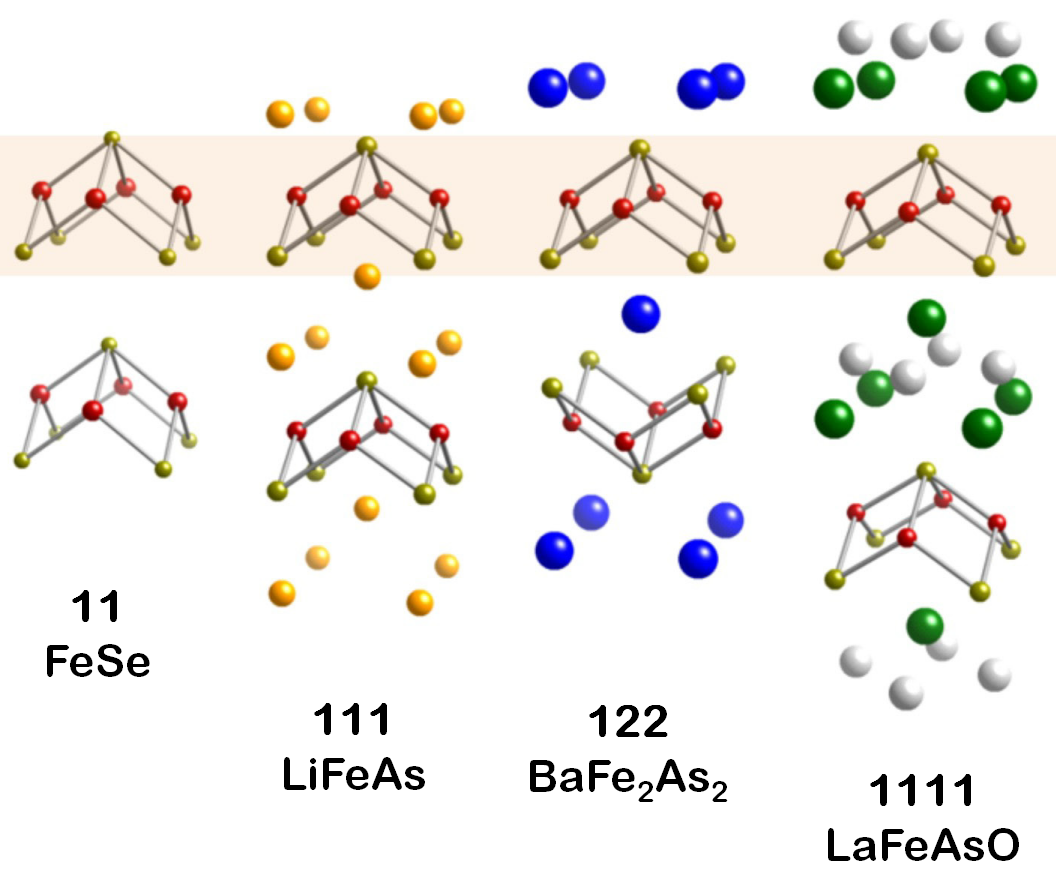}
\caption{Crystal structures of some of iron based superconductors, after \cite{PaglioneNPh2010}.
\label{FeSC_structures}}
\end{center}
\end{figure}

\begin{figure*}
\begin{center}
\includegraphics[width=0.8\textwidth]{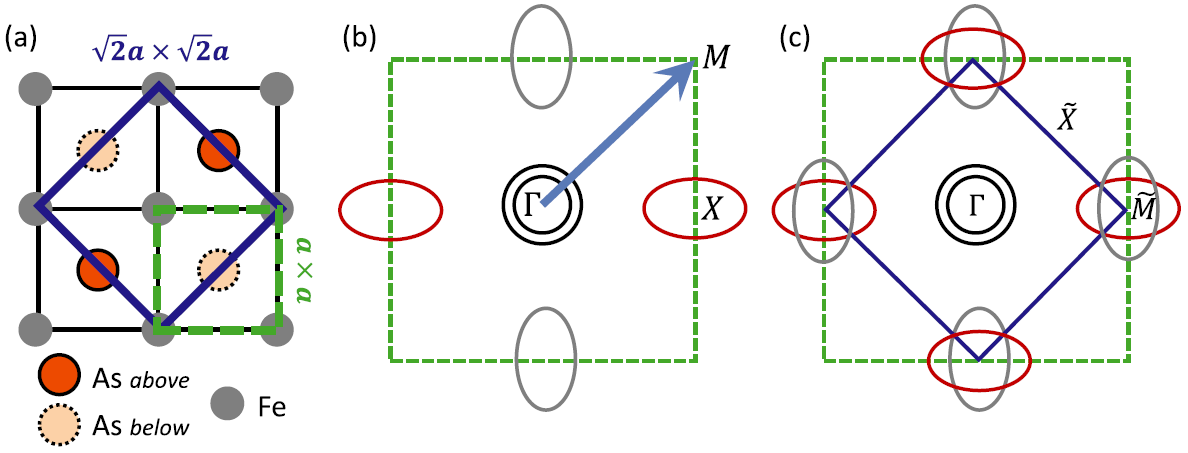}
\caption{(a) FeAs lattice indicating As above and below the Fe plane. Dashed green and solid blue squares indicate 1- and 2-Fe unit cells, respectively. (b) Schematic 2D Fermi surface in the 1-Fe BZ whose boundaries are indicated by a green dashed square. (c) Fermi sheets in the folded BZ whose boundaries are now shown by a solid blue square. After \cite{HirschfeldRPP2011}.
\label{FeAs_FS}}
\end{center}
\end{figure*}

To date, there is a number of useful and comprehensive reviews on the diverse properties of FeSC \cite{SadovskiiPU2008, IvanovskiiPU2008, IzyumovPU2008, IshidaJPSJ2009, JohnstonAPh2010, PaglioneNPh2010, WenARoCMP2011} and on the pairing models \cite{WenARoCMP2011, HirschfeldRPP2011, ChubukovAR2012}. The scope of this review is smaller but twofold. On one hand, I want to give a simple, even oversimplified introduction to the FeSC physics. On the other hand, I want to advocate an opinion that all the complexity of FeSC properties are encapsulated in their complex but well defined and rather common multi-band electronic structure. For many compounds, this structure has been determined in numerous angle resolved photoemission experiments (ARPES), and one of the scopes of this review is to show that while the overall agreement between the measured and calculated band structures is very good, it is the observed small differences \cite{KordyukFPS2011} that may help to understand the mechanisms of the magnetic ordering and superconducting pairing in FeSC.

\begin{figure*}
\begin{center}
\includegraphics[width=0.9\textwidth]{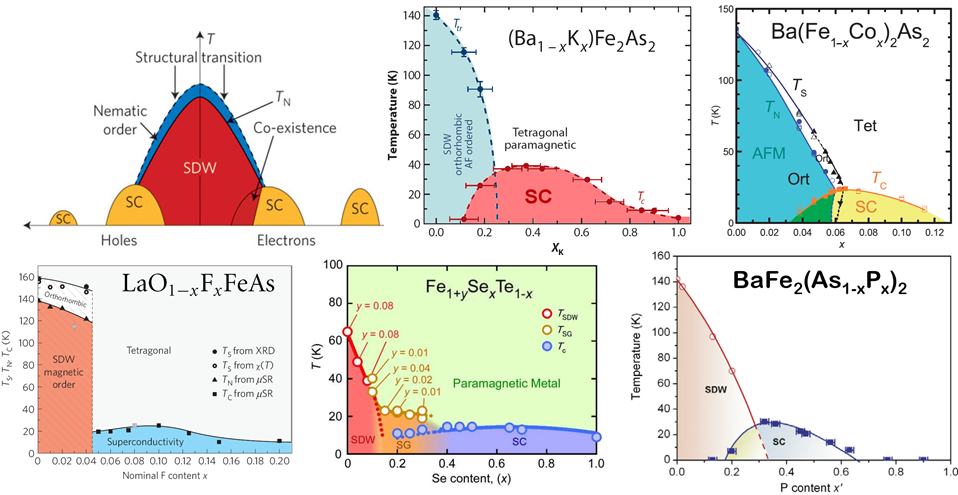}
\caption{Examples of the FeSC phase diagrams: a schematic one \cite{ChubukovAR2012}, and the diagrams measured for (Ba$_{1-x}$K$_{x}$)Fe$_2$As$_2$ \cite{WenARoCMP2011},  Ba(Fe$_{1-x}$Co$_{x}$)$_2$As$_2$ \cite{NandiPRL2010}, La(O$_{1-x}$F$_{x}$)FeAs \cite{LuetkensNM2009}, Fe$_{1+y}$Se$_{x}$Te$_{1-x}$ \cite{Katayama2010}, BaFe$_2$(As$_{1-x}$P$_{x}$)$_2$ \cite{JiangJPCM2009}.
\label{FeSC_phase_diagrams}}
\end{center}
\end{figure*}

\section{Iron based superconductors}

There are many families of FeSC with different structure and composition already known \cite{SadovskiiPU2008, IvanovskiiPU2008, IzyumovPU2008, IshidaJPSJ2009, JohnstonAPh2010, PaglioneNPh2010, WenARoCMP2011} but all share a common iron-pnictogen (P, As) or iron-chalcogen planes (Se, Te), as shown in Fig.\,\ref{FeSC_structures} \cite{PaglioneNPh2010}. All the compounds share similar electronic band structure in which the electronic states at the Fermi level are occupied predominantly by the Fe $3d$ electrons. The structure itself is quite complex and, in most cases, consist of five conduction bands that result in rather complex Fermiology that changes rapidly with doping and, consequently, leads to many unusual superconducting and normal state properties. Fig.\,\ref{FeAs_FS} shows the square FeAs lattice and the corresponding Fermi surface for a stichometric parent compound.   Fig.\,\ref{FeSC_phase_diagrams} provides examples of the FeSC phase diagrams with distinct areas of the antiferromagnetically ordered spin density wave (marked as AFM or SDW and bordered by the N\'{e}el temperature $T_N$) and superconducting (SC, $T_c$) phases, reminding the extensively discussed phase diagram of CuSC \cite{KordyukLTP2006}. Here I briefly review some of the most interesting and most studied FeSC materials with the references to their properties and experimental (ARPES) studies of their electronic structure, which will be important for the following discussion.

\textbf{1111.}
Starting from LaO$_{1-x}$F$_x$FeAs \cite{KamiharaJACS2008}, the 1111 family keeps the records of $T_c$: NdFeAsO$_{1-y}$ (54 K), SmFeAsO$_{1-x}$F$_{x}$ (55 K), Gd$_{0.8}$Th$_{0.2}$FeAsO (56.3 K), but the material is hard to study. First, the available single crystals are too small---for all members of the family they grow as thin platelets up to $200 \times 200 \times 10$ $\mu \textrm{m}^3$ only \cite{MollNM2010}. Second, the termination of the crystal reveals a polar surface with distinct surface states that are markedly different from the bulk electronic structure \cite{EschrigPRB2010} and highly complicate the use of any surface sensitive experimental probe such as ARPES \cite{LiuPRB2010}.

\textbf{122.}
The 122 family consists of a variety of different compounds with wide ranges of doping in both hole and electron sides \cite{WenARoCMP2011} that form a rich phase diagram (see Fig.\,\ref{FeSC_phase_diagrams}) where the superconductivity and magnetism compete or coexist. The most studied compounds are the hole doped Ba$_{1-x}$K$_{x}$Fe$_2$As$_2$ (BKFA) with $T_c^\textrm{max} =$ 38 K \cite{RotterPRL2008} and the electron doped Ba(Fe$_{1-x}$Co$_{x}$)$_2$As$_2$ (BFCA), 22 K \cite{SefatPRL2008, NiPRB2008}. Both share the same parent compound,  BaFe$_2$As$_2$ (BFA), which is a compensated metal, i.e. the total volume of its three hole Fermi surfaces (FS's) is equal to the total volume of two electron FS's \cite{BrouetPRB2009, KondoPRB2010}. BFA goes into magnetically ordered phase below 140 K \cite{RotterPRB2008} and never superconducts. An extremely overdoped BKFA is a stoichiometric KFe$_2$As$_2$ (KFA) \cite{SatoPRL2009}, which is non-magnetic, with $T_c =$ 3 K. There is also an interesting case of isovalent dopping, BaFe$_2$(As$_{1-x}$P$_{x}$)$_2$ (BFAP)  ($T_c =$ 30 K) \cite{JiangJPCM2009} with similar phase diagram (see Fig.\,\ref{FeSC_phase_diagrams}).

To this, one can add a number of similar compounds: Ba$_{1-x}$Na$_{x}$Fe$_2$As$_2$ (BNFA) ($T_c^\textrm{max} =$ 34 K) \cite{CortesGil2010, Aswartham2012}, Ca$_{1-x}$Na$_{x}$Fe$_2$As$_2$ ($\sim$20 K) \cite{WuJPCM2008}, CaFe$_2$As$_2$ ($T_N =$ 170 K, $T_c >$ 10 K under pressure) \cite{ParkJPCM2008}, EuFe$_2$(As$_{1-x}$P$_{x}$)$_2$  ($T_c =$ 26 K) \cite{RenPRL2009}, etc. \cite{JohnstonAPh2010}

As a consequence of good crystal quality and variety of compounds, the 122 family is the most studied by ARPES \cite{RichardRPP2011} (see \cite{LiuPRL2008, DingEPL2008, ZabolotnyyN2009, EvtushinskyPRB2009, RichardPRL2009, EvtushinskyNJP2009, EvtushinskyJPSJ2011} for BKFA, \cite{BrouetPRB2009, ThirupathaiahPRB2010, LiuNP2010, LiuPRB2011, YiPNAS2011, HeumenPRL2011} for BFCA, \cite{YiPRB2009, FinkPRB2009, KondoPRB2010, RichardPRL2010, KimPRB2011} for BFA, \cite{SatoPRL2009, YoshidaJPCS2011} for KFA, \cite{LiuPRL2009, KondoPRB2010} for CaFe$_2$As$_2$, \cite{YoshidaPRL2011, ZhangNP2012} for BFAP, \cite{ThirupathaiahPRB2011} for EuFe$_2$(As$_{1-x}$P$_{x}$)$_2$). The ARPES spectra well represent the bulk electronic structure of this family, at least, for the hole doped BKFA, BNFA, and BFAP, where the superconducting gap is routinely observed \cite{DingEPL2008, EvtushinskyPRB2009, Evtushinsky2011} and is in a good agreement with the bulk probes \cite{EvtushinskyNJP2009, EvtushinskyJPSJ2011}. This poses the 122 family as the main arena to study the rich physics of the iron-based superconductors.

\begin{figure*}
\begin{center}
\includegraphics[width=0.8\textwidth]{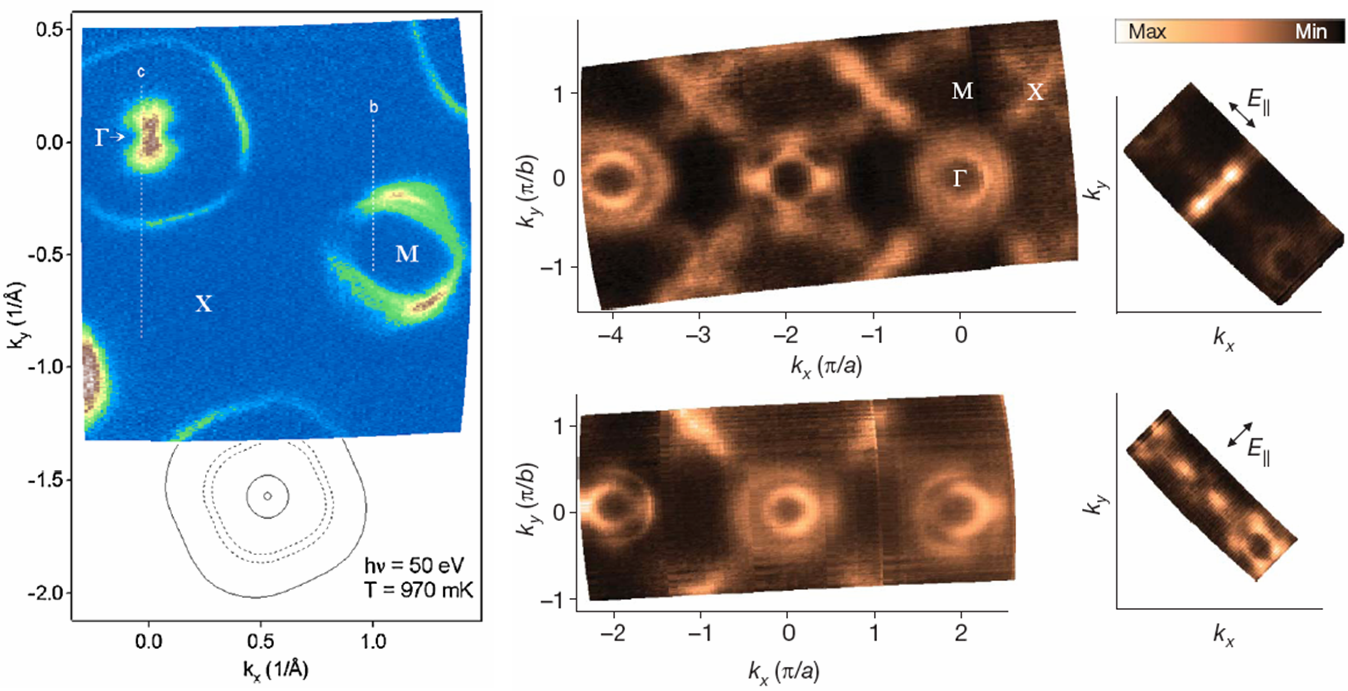}
\caption{Fermi surface (FS) maps measured by ARPES for LiFeAs \cite{BorisenkoPRL2010} (left) and an optimally doped Ba$_{1-x}$K$_{x}$Fe$_2$As$_2$ (BKFA) \cite{ZabolotnyyN2009} (right).
\label{FS_Maps}}
\end{center}
\end{figure*}

\textbf{111.} Being highly reactive with air and, consequently, more challenging to study, the 111 family has gave many interesting results that keep growing. The main representative of the family, LiFeAs \cite{WangSSC2008, TappPRB2008}, is the most ``arpesable'' compound \cite{BorisenkoPRL2010, KordyukPRB2011, BorisenkoSym2012}. It grows in good quality single crystals \cite{MorozovCGD2010} that cleave between the two Li layers, thus revealing a non-polar surface with protected topmost FeAs layer; it is stoichiometric, i.e. impurity clean; it has the transition temperature about 18\,K and one can measure the superconducting gap by ARPES and compare its value to bulk techniques; it is non-magnetic and, consequently, the observed band structure is free of SDW replicas; and, finally, its electronic bands are the most separated from each other that allows one to disentangle them most easily and analyze their fine structure \cite{KordyukPRB2011}. Fig.\,\ref{FS_Maps} shows the FS maps measured by ARPES for LiFeAs (left) and an optimally doped BKFA (right).

NaFeAs is another member of 111 family. It shows three successive phase transitions at around 52, 41, and 23 K, which correspond to structural, magnetic, and superconducting transitions, respectively \cite{ChenPRL2009, LiPRB2009}. The compound is less reactive with the environment than LiFeAs but the exposure to air strongly affects $T_c$ \cite{TanatarPRB2012}. Replacing Fe by either Co or Ni suppresses the magnetism and enhances superconductivity \cite{ParkerPRL2010}. For ARPES on NaFeAs, see \cite{HePRL2010, HeJPCS2011}.

\textbf{11.}
The binary FeAs does not crystallize in the FeAs-layered structure (it adopts an orthorhombic structure consisting of distorted FeAs$_6$ octahedra unlike the superconducting ferro-pnictides in which FeAs$_4$ tetrahedra form square lattices of iron atoms \cite{SegawaJPSJ2009}), but the FeSe does. So, the 11 family is presented by simplest ferro-chalcogenides FeSe and FeTe, and their ternary combination  FeSe$_{x}$Te$_{1-x}$ \cite{SalesPRB2009}. The FeSe has been found to superconduct at approximately 8 K \cite{HsuPNAS2008}), and up to 37 K under pressure \cite{ImaiPRL2009}. Fe$_{1+y}$Se$_{x}$Te$_{1-x}$ shows maximum $T_c$ about 14 K at $x = 0.5$ \cite{SalesPRB2009, Katayama2010, FedorchenkoLTP2011}. The crystals grow with excess ($y$) Fe atoms that present beyond those needed to fill the Fe square lattice layers and go into interstitial positions within the Te layers \cite{JohnstonAPh2010}. For ARPES on 11 family, see \cite{XiaPRL2009, TamaiPRL2010, NakayamaPRL2010, MiaoPRB2012}.

\textbf{245 or x22.}
The attempts to intercalate FeSe, the simplest FeSC, resulted in discovery of a new family A$_x$Fe$_{2-y}$Se$_2$ (A stands for alkali metal: K, Rb, Cs, Tl) with $T_c$ up to 30 K and with exceptionally high N\'{e}el temperature ($>$500 K) and magnetic moment ($>$$3 \mu_B$) \cite{GuoPRB2010, WangNC2011, LiPRB2011}. This family called most often ``245" because of its parent compound A$_{0.8}$Fe$_{1.6}$Se$_2$ $\equiv$ A$_2$Fe$_4$Se$_5$. It is interesting that their resistivity shows insulating behavior down to 100 K and superconductivity seems to occur from an antiferromagnetic semiconductor \cite{YanSR2012}. This, however, is not consistent with ARPES results which show the presence of a Fermi surface \cite{QianPRL2011, MouPRL2011, ZhangNM2011, ChenPRX2011}. It is even more interesting that the observed FS is completely electron-like that, seemingly, contradicts to the most popular $s\pm$ scenario for superconducting pairing \cite{WenARoCMP2011, HirschfeldRPP2011, ChubukovAR2012}. Recently, it has been shown \cite{BorisenkoXXX2012}, that the puzzling behavior of these materials is the result of separation into metallic and antiferromagnetic insulating phases, from which only the former becomes superconducting, while the later has hardly any relation to superconductivity. The superconducting phase has an electron doped composition A$_x$Fe$_2$Se$_2$ (so, the family can be called ``x22"). Similar conclusion has been made based on neutron scattering experiments \cite{FriemelPRB2012}.

\section{Magnetism}

Naturally, the magnetic properties of FeSC are very rich and far from be completely understood \cite{LumsdenJPCM2010}, but since the focus of this review is in superconductivity, I will discuss only two issues: coexistence of static magnetism and superconductivity and role of spin fluctuations.

\subsection{Magnetic ordering}

\begin{figure}
\begin{center}
\includegraphics[width=0.45\textwidth]{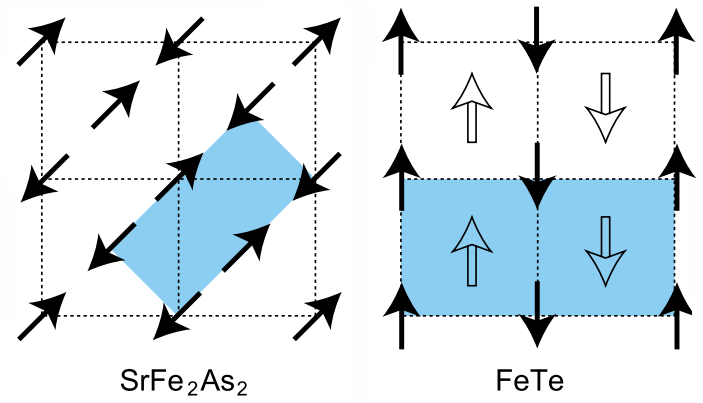}
\caption{In-plane magnetic structure common for the 1111 and 122 parent compounds (left) and for parent 11 compound (FeTe, right). The shaded areas indicate the magnetic unit cells. After \cite{Li2009FeTe}.
\label{SDW}}
\end{center}
\end{figure}

Nearly perfect FS nesting in many parent compounds (which are compensated metals) suggests us to expect some static density wave with the nesting vector ($\pi,\pi$), as a way to lower the kinetic energy of the electrons (Peierls transition) \cite{GrunerRMP1988}. Therefore, the realization of the antiferromagnetic spin density wave in those compounds is quite natural---the most easy ordering for Fe lattice is the spin ordering \cite{DongEPL2008}. Indeed, almost all parent compounds enters the antiferromagnetic SDW below N\'{e}el temperature with exactly the same wavevector. Such a most common spin configuration on Fe atoms is shown in Fig.\,\ref{SDW} (left) \cite{Li2009FeTe}. This said, there are different opinions based on importance of interaction of the localized spins \cite{YildirimPRL2008, MazinPhC2009, YinPRL2010} (see also \cite{LumsdenJPCM2010} for review on this topic). From experiment, there are both pro and con arguments on this problem. Pro: any time when the FS nesting is good (BFA \cite{YiPRB2009, KondoPRB2010, RichardPRL2010} and other parent compounds of 122 family \cite{YoshidaPRL2011, ZhangNP2012, ThirupathaiahPRB2011}, NaFeAs \cite{HePRL2010, HeJPCS2011}), the SDW is present, and when nesting is poor or absent (superconducting BKFA \cite{LiuPRL2008, DingEPL2008}, BFCA \cite{BrouetPRB2009, ThirupathaiahPRB2010}, BFAP \cite{YoshidaPRL2011}, and stoichiometric LiFeAs \cite{BorisenkoPRL2010}), there is no magnetic ordering. Con: Fe$_{1+y}$Te shows different spin order, see Fig.\,\ref{SDW} (right) \cite{Li2009FeTe}, despite having very similar FS topology (as it follows from calculations \cite{SubediPRB2008} and ARPES study \cite{XiaPRL2009}). So, one may conclude that the mechanism of the magnetic ordering in FeAS is not yet clear, but for the scope of this review it is important to know that this ordering is routinely observed in many compounds, \emph{always} neighboring the superconducting phase and often coexisting with it.

Since static magnetism and superconductivity coexist on the phase diagrams for a number of FeSC \cite{StewartRMP2011}, it is important to answer the questions: (1) do they coexist microscopically and (2) do magnetism and superconductivity evolve from the same conduction electrons? The latter is related to the ``itinerant \emph{vs.} localized" problem and was briefly discussed above. The problem of coexistence on the microscopic scale is related to sample homogeneity and has been addressed in a number of publications (see \cite{StewartRMP2011} for a short review). In particular, for BFCA crystals, the homogeneity of superconducting state was demonstrated by magneto-optic imaging \cite{ProzorovPhC2009} down to 2 $\mu$m and by NMR \cite{JulienEPL2009} down to the sub-nanometer scale. Another 122 compound, BKFA, is known to be inhomogeneous \cite{EvtushinskyPRB2009}, and some separation of the magnetic and superconducting regions has been found on a nanometer scale \cite{ParkPRL2009}. An evidence for homogeneity has been reported for one of 245 family, K$_{0.8}$Fe$_{1.6}$Se$_2$, \cite{ShermadiniPRL2011}, but not confirmed by magnetic measurements on similar samples \cite{ShenEPL2011}. Clear phase separation in other, Rb based 245, has been recently demostrated by ARPES \cite{BorisenkoXXX2012} and by inelastic neutron scattering (INS) \cite{FriemelPRB2012}. Also, it has been shown, that in EuFe$_2$As$_2$ under pressure \cite{KuritaPRB2011}, similarly to the quaternary borocarbides \cite{GuptaAP2006}, the antiferromagnetism is realized on the Eu sublattice, affecting the superconductivity on the Fe sublattice. So, one may conclude that while on some systems like 245 the magnetic and superconducting phases are spatially separated, the question of coexistence in other FeSC systems requires more careful study.

The neighboring is close and interesting issue. FeSC are perfect systems for realization of the CDW (or SDW) induced superconductivity, the idea which had been suggested  long ago \cite{KopaevZETF1970, MattisPRL1970, RusinovJETP1973} and widely discussed \cite{BalseiroPRB1979, KopaevPLA1987, GabovichLTP2000, GabovichACMP2010}. For a slightly non-stoichiometric system, the band gap cannot kill the FS completely since some extra carriers should form small FS pockets and place the Van Hove singularity (vHs) close to the Fermi level. This mechanism is supported empirically since there are many known systems where superconductivity occurs at the edge of CDW or SDW phase \cite{FangPRB2005, MorosanNP2006, Kato1988}. On the other hand, the related increase of the density of states seems to be too small to explain the observed $T_c$'s within the standard BCS model. I this sense, the conclusion of this review about importance of the proximity of FS to Lifshitz transition for superconductivity can help to understand the density wave induced superconductivity, in general.

\subsection{Spin-fluctuations}

If a magnetically mediated pairing mechanism takes place in FeSC, the spin-fluctuation spectrum must contain the necessary spectral weight to facilitate pairing \cite{LumsdenJPCM2010}. It is also expected that fingerprints of its structure will be recognizable in one-particle spectral function, like in case of cuprates (for example, see \cite{DahmNP2009, KordyukEPJ2010}).

The spin dynamics in FeSC is revealed primarily by INS and, in some cases, supplemented by NMR measurements (see \cite{LumsdenJPCM2010} for review).

First, the correlation between the spectral weight of the spin-fluctuations and superconductivity is observed. In at least two cases (BFCA \cite{NingPRL2010, MatanPRB2010} and LaFeAsO$_{1-x}$F$_{x}$ \cite{WakimotoJPSJ2010}), when antiferromagnetically ordered parent compounds are overdoped by electron doping, the spin fluctuations vanish together with the FS hole pocket \cite{BrouetPRB2009} and superconductivity. This is compatible with the idea that the spin fluctuations are completely defined by the electronic band structure and play important role in superconductivity.

Second, the correlation between the normal state spin excitations and electronic structure is found to be common for all FeSC \cite{LumsdenJPCM2010}. In particurar, even in Fe$_{1+y}$Se$_{x}$Te$_{1-x}$ \cite{IikuboJPSJ2009}, an interesting early development in the study of the spin excitations was that, in contrast to the parent FeTe, the spin fluctuations in superconducting samples were found at a similar wavevector as found in the other Fe-based materials. Also, there is another common feature, a quartet of low energy incommensurate inelastic peaks characterized by the square lattice wavevectors ($\pi \pm \xi$ ,$\pi$) and ($\pi$, $\pi \pm \xi$), observed for BFCA \cite{LesterPRB2010, LiPRB2010}, Fe$_{1+y}$Se$_{x}$Te$_{1-x}$ \cite{ArgyriouPRB2010}, and ${\text{CaFe}}_{2}{\text{As}}_{2}$ \cite{DialloPRB2010}, in analogy to CuSC \cite{VignolleNP2007}.

Third, the ``resonance peak" in the spin-fluctuation spectrum has been observed in many FeSC compounds in superconducting state, that is considered by many authors as an evidence for a sign change of the superconducting order parameter \cite{LumsdenJPCM2010, StewartRMP2011}.

The spin resonance, the resonance in the dynamic spin susceptibility, occurs indeed because of its divergence through a sign change of the superconducting order parameter on different parts of the Fermi surface \cite{MonthouxPRL1994}. In cuprates, it was associated with the ``resonance peak", observed in INS experiments, and considered as one of the arguments for $d$-wave symmetry of the superconducting gap. In FeSC, the resonance peak was predicted to be the most pronounced for the $s\pm$ gap \cite{KorshunovPRB2008, MaierPRB2008} and, indeed, the peaks in INS spectra had been observed for a number of compounds: BKFA \cite{ChristiansonN2008}, BFCA \cite{LumsdenPRL2009, InosovNP2010}, Fe$_{1+y}$Se$_{x}$Te$_{1-x}$ \cite{ArgyriouPRB2010, LeePRB2010}, ${\mathrm{Rb}}_{2}{\mathrm{Fe}}_{4}{\mathrm{Se}}_{5}$ \cite{ParkPRL2011, FriemelPRB2012}, etc.\,\cite{LumsdenJPCM2010}. However, one should realize, that the peak in the dynamic susceptibility is not necessarily caused by the spin resonance but can be due to a peak in the bare susceptibility (Lindhard function), which, as a result of self-correlation of electronic Green's function, is expected to be peaked in energy at about $2\Delta$ and in momentum at the FS nesting vectors \cite{InosovPRB2007}. In \cite{OnariPRB2010}, in contrast to \cite{MaierPRB2008}, it has been shown that a prominent hump structure appears just above the spectral gap by taking into account the quasiparticle damping in SC state. The obtained hump structure looks similar to the resonance peak in the $s\pm$-wave state, although the height and the weight of the peak in the latter state is much larger. This shows that in order to support the sign charge scenario, not only the presence of the peak in INS spectra but also its spectral weight should be considered. The later is not trivial task. In Ref.\,\onlinecite{InosovNP2010}, for example, the INS measurements were calibrated in the absolute scale and the spectral weight of the resonance in BFCA has been found to be comparable to ones in cuprates.

In summary, the spin-fluctuation spectra in FeSC, looks, at first glance, similar to the ones in CuSC in terms of appearance and correlation with electronic structure, but its accurate interpretation requires more efforts. As very last example for this, a combined analysis of neutron scattering and photoemission measurements on superconducting FeSe$_{0.5}$Te$_{0.5}$ \cite{LeePRB2010} has shown that while the spin resonance occurs at an incommensurate wave vector compatible with nesting, neither spin-wave nor FS nesting models can describe the magnetic dispersion. The authors propose that a coupling of spin and orbital correlations is key to explaining this behavior.

\begin{figure*}
\begin{center}
\includegraphics[width=1\textwidth]{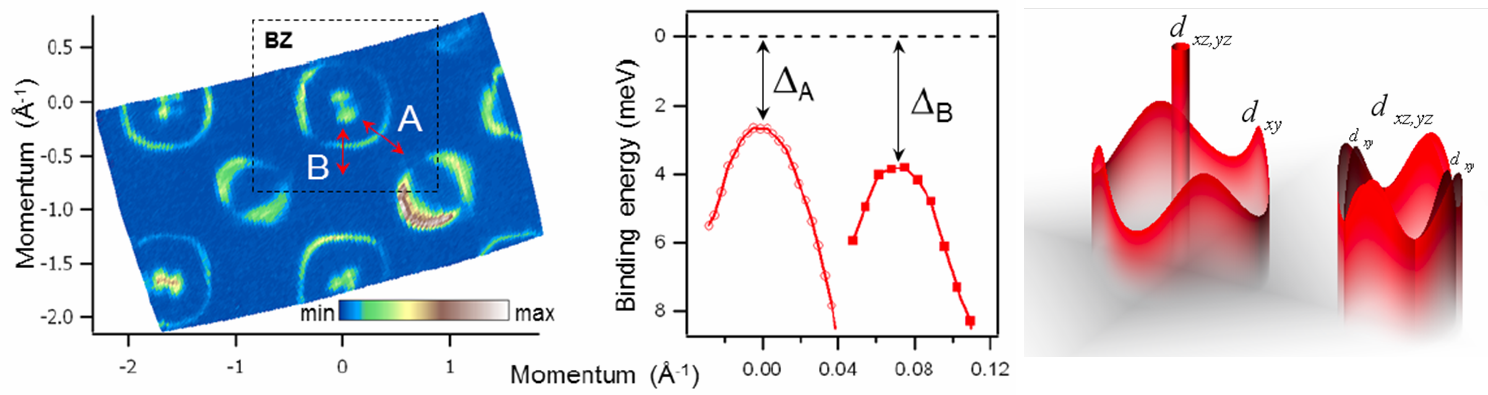}
\caption{Superconducting gap symmetry in LiFeAs. Experimental Fermi surface (left). The experimental dispersions (center) measured along the cuts A and B. A sketch of distribution of the superconducting gap magnitude over Fermi surfaces (right). After \cite{HirschfeldRPP2011}.
\label{LFA_Gaps}}
\end{center}
\end{figure*}

\subsection{Pseudogap}

Surprisingly, the pseudogap in FeSC is not a hot topic like in cuprates \cite{TimuskRPP1999}. From a nearly perfect FS nesting one would expect the pseudogap due to incommensurate ordering like in transition metal dichalcogenides \cite{BorisenkoPRL2008} and, may be, in cuprates \cite{KordyukPRB2009}. If the pseudogap in cuprates is due to superconducting fluctuations \cite{Huefner2008}, then it would be also natural to expect it in FeSC.

In NMR data, the decrease in $1/T_1T$ in some of 1111 compounds and BFCA \cite{IshidaJPSJ2009} was associated with the pseudogap. The interplane resistivity data for BFCA over a broad doping range also shows a clear correlation with the NMR Knight shift, assigned to the formation of the pseudogap \cite{TanatarPRB2010}. In SmFeAsO$_{1-x}$, the pseudogap was determined from resistivity measurements \cite{SolovevLTP2009}. The evidence for the superconducting pairs in the normal state (up to temperature $T \approx 1.3 T_c$) has been obtained using point-contact spectroscopy on BFCA film.

An evidence for the pseudogap has been reported from photoemission experiments on polycrystalline samples (e.g., see \cite{SatoJPSJ2008, HaiYun2008}) and in some ARPES experiments on single crystals \cite{XuNC2011}, but this is supported neither by other numerous ARPES studies \cite{EvtushinskyPRB2009, RichardPRL2009, Evtushinsky2011, ShimojimaSci2011, BorisenkoSym2012, ZhangNP2012, Evtushinsky2012} nor by STM measurements \cite{YinPhC2009, MasseeEPL2010}. The absence of the pseudogap in ARPES spectra may be just a consequence of low spectral weight modulation by the magnetic ordering that may question its importance for superconductivity, discussed in previous section.

\section{Superconductivity}

In 1111 and 122 systems, superconductivity emerges upon electron or hole doping, or can be induced by pressure \cite{SefatRPP2011} or by isovalent doping. In 111 systems, superconductivity emerges already at zero doping instead of  magnetic order (in LiFeAs) or together with it (in NaFeAs). There are several important experimentally established tendencies, which are followed by many representatives of iron-based family with highest $T_c$ \cite{Evtushinsky2012}: large difference in superconducting gap magnitude on different FS pockets \cite{DingEPL2008, EvtushinskyNJP2009, HardyEPL2010, PopovichPRL2010}, $\Delta/T_c$ value, that is similar to cuprates and much higher than expected from BCS \cite{HardyEPL2010, EvtushinskyNJP2009, Evtushinsky2011}, correlation of $T_{\rm \text{c}}$ with anion height \cite{OkabePRB2010}. The complexity of the electronic structure of FeSC was originally an obstacle on the way to its understanding \cite{DingEPL2008, ZabolotnyyN2009}, but at closer look, such a variety of electronic states turned out to be extremely useful for uncovering the correlation between orbital character and pairing strength \cite{Evtushinsky2012} and, more general, between electronic structure and superconductivity \cite{KordyukFPS2011}.

In this section I briefly discuss the existent pairing models, the experimental, mainly ARPES data on superconducting gap symmetry, and the observed general correlation of the electronic band structure with $T_c$.

\subsection{Pairing models}

From similarity of the phase diagrams for FeSC and cuprates, it was proposed that the pairing in FeSC is also mediated by spin-fluctuations that assumes the sign change of the superconducting order parameter. Then, to adopt the FS geometry of FeSC, the symmetry of the sign change should be different from $d$-wave symmetry of cuprates and can be satisfied by an extended s-wave pairing with a sign reversal of the order parameter between different Fermi surface pockets \cite{MazinPRL2008}. Today, the most of researchers do believe that the gap does have $s\pm$ symmetry, at least in weakly and optimally doped FeSCs (see recent reviews \cite{HirschfeldRPP2011, ChubukovAR2012}).

This said, numerous studies of superconductivity in FeSCs demonstrated that the physics of the pairing could be more involved than it was originally
thought because of the multiorbital/multiband nature of low-energy electronic excitations \cite{ChubukovAR2012}. It turns out that both the symmetry and the structure of the pairing gap result from rather nontrivial interplay between spin-fluctuation exchange, Coulomb repulsion, and the momentum structure of the interactions. In particular, an $s\pm$-wave gap can be with or without nodes, depending on the orbital content of low-energy excitations, and can even evolve into a $d$-wave gap with hole or electron overdoping. In addition to spin fluctuations, FeSCs also possess charge fluctuations that can be strongly enhanced \cite{OnariPRL2009, YinPRL2010} due to proximity to a transition into a state with an orbital order. This interaction can give rise to a conventional $s$-wave pairing.

The experimental data on superconductivity show very rich behavior, superconducting gap structures appear to vary substantially from family to family, and even within families as a function of doping or pressure \cite{HirschfeldRPP2011}. The variety of different pairing states raises the issue of whether the physics of FeSCs is model dependent or is universal, governed by a single underlying pairing mechanism \cite{ChubukovAR2012}.

In favor of $s\pm$ symmetry, there are natural expectation that spin-fluctuations mediate pairing in FeSC, the observation of spin resonances by INS, which implies the sign change of $\Delta$ as discussed above, and numerous experimental evidences for the nodal gap \cite{NakaiPRB2010, YamashitaPRB2011, ZhangNP2012} (see also references in \cite{HirschfeldRPP2011, ChubukovAR2012}). It was also argued \cite{VorontsovPRB2010} that the very presence of the coexistence region between SC and stripe magnetism in FeSCs is a fingerprint of an $s\pm$ gap, because for an $s^{++}$ gap a first-order transition between a pure magnetic and a pure SC state is much more likely \cite{ChubukovAR2012}.

On the other hand, several cons come from ARPES. There is an evidence for strong electron-phonon coupling in LiFeAs \cite{KordyukPRB2011, BorisenkoSym2012}. The accurately measured gap anisotropy is difficult to reconcile with the existent $s\pm$ models but with $s^{++}$ models based on orbital fluctuations assisted by phonons \cite{OnariPRL2009, KontaniPRL2010, YanagiPRB2010}. The remnant superconductivity in KFe$_2$As$_2$, and, actually, for all overdoped BKFA started from the optimally doped one \cite{ZabolotnyyN2009}, should have different symmetry since only hole like FSs are present \cite{KordyukFPS2011}. The same is applicable for A$_x$Fe$_{2-y}$Se$_2$ where only electron-like FSs are present \cite{QianPRL2011, MouPRL2011, ZhangNM2011, ChenPRX2011, BorisenkoXXX2012}.

In \cite{ChubukovAR2012} it was suggested that in both A$_x$Fe$_{2-y}$Se$_2$ and KFe$_2$As$_2$ cases the gap symmetry may be $d$-wave, though with different nodes. In \cite{Khodas2012} it is argued that $s\pm$ symmetry in A$_x$Fe$_{2-y}$Se$_2$ can be realized due to inter-pocket pairing, i.e.~$\Delta$ changes sign between electron pockets. Another possibility \cite{HirschfeldRPP2011} for the order parameter to change sign in A$_x$Fe$_{2-y}$Se$_2$, is taking into account the finite energy of the coupling boson that should be higher than the binding energy of the top of the hole band in $\Gamma$-point, but one can hardly describe rather high $T_c$ in 245 family within such a mechanism.

\subsection{Superconducting gap}

The best FeSC for ARPES and, consequently, the systems on which the most reliable data on superconducting gap can be obtained, are LiFeAs, BKFA (and similar hole doped compounds), and BFAP.

\begin{figure*}
\begin{center}
\includegraphics[width=0.8\textwidth]{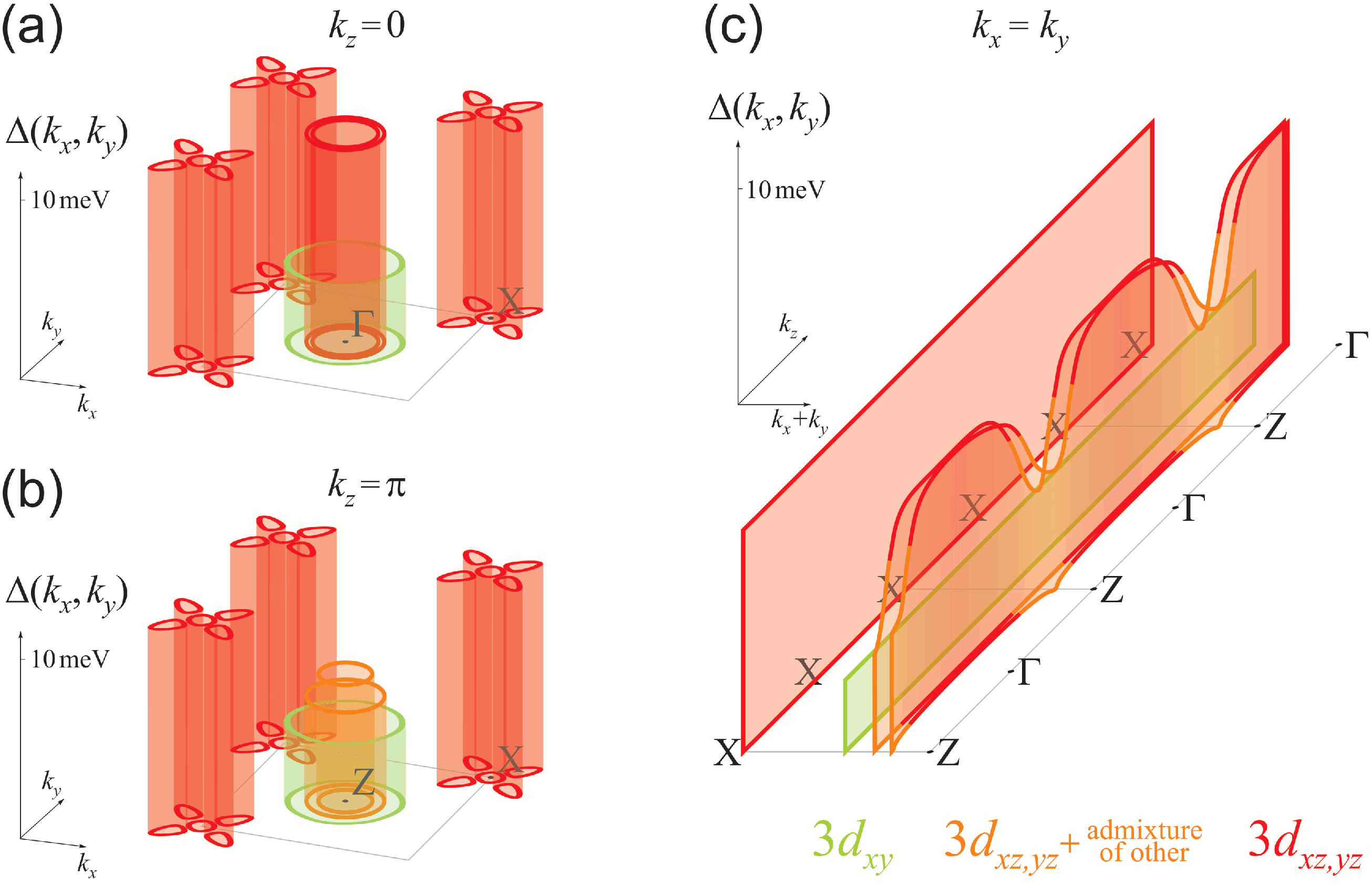}
\caption{
Three-dimensional distribution of the superconducting gap and orbital composition of the electronic states at the Fermi level of Ba$_{1-x}$K$_{x}$Fe$_2$As$_2$ (BKFA). (a) Distribution of the superconducting gap (plotted as height) and distribution of the orbital composition for the states at the Fermi level (shown in color: $d_{xz,yz}$\,---\,red, $d_{xy}$\,---\,green, $d_{xz,yz}$ with admixture of other orbitals\,---\,orange) as function of $k_x$ and $k_y$ at constant $k_z=0$; (b) the same, only for $k_z=\pi$; (c) same distributions as function of in-plane momentum, directed along BZ diagonal, and $k_z$. Note unambiguous correlation between the color and height, i.e. there is strong correlation between the orbital composition and superconducting gap magnitude. After \cite{Evtushinsky2012}.
\label{3D_gap}}
\end{center}
\end{figure*}

LiFeAs allows the most careful determination of the gap value \cite{BorisenkoPRL2010, BorisenkoSym2012}. Accurate measurements at 1 K have allowed to detect the variations of $\Delta$ over the FS with relative precision of 0.3 meV and the result is the following \cite{BorisenkoSym2012} (see Fig.\,\ref{LFA_Gaps}): On the small hole-like FS at $\Gamma$-point of $d_{xz/yz}$ origin, that, at some $k_z$, only touches the Fermi level, the largest superconducting energy gap of the size of 6 meV opens and is in agreement with tunneling spectroscopy \cite{HankePRL2012}. Along the large 2D hole-like FS of $d_{xy}$ character the gap varies around 3.4 meV roughly as 0.5 meV $\cos (4\phi)$, being minimal at the direction towards the electron-like FS. The gap on the outer electron pocket is smaller than on the inner one and both vary around 3.6 meV as 0.5 meV $\cos (4\phi)$, having maximal values at the direction towards $\Gamma$-point. The detected gap anisotropy is difficult to reconcile with coupling through spin fluctuations and the sign change of the order parameter but fits better to the model of orbital fluctuations assisted by phonons \cite{OnariPRL2009, KontaniPRL2010, YanagiPRB2010}.

In BKFA, the superconducting gap was studied by means of various experimental techniques \cite{EvtushinskyNJP2009, PopovichPRL2010}, and vast majority of the results can be interpreted in terms of presence of comparable amount of electronic states gapped with a large gap ($\Delta_{\rm large}=$10--11\,meV) and with a small gap ($\Delta_{\rm small}<4$\,meV). The in-plane momentum dependence of the superconducting gap, determined in early ARPES studies, is the following: the large gap is located on all parts of the FS except for the outer hole-like FS sheet around $\Gamma$-point \cite{DingEPL2008, EvtushinskyPRB2009}. In \cite{Evtushinsky2012}, a clear correlation between the orbital character of the electronic states and their propensity to superconductivity is observed in hole-doped BaFe$_2$As$_2$: the magnitude of the superconducting gap maximizes at 10.5\,meV exclusively for iron $3d_{xz,yz}$ orbitals, while for others drops to 3.5\,meV (see Fig.\,\ref{3D_gap}).

In BFAP, motivated by earlier reported evidences for the nodal gap from NMR \cite{NakaiPRB2010} and angle-resolved thermal conductivity \cite{YamashitaPRB2011}, the superconducting gap was measured by ARPES \cite{ZhangNP2012} as function of $k_z$, the out-of-plane momentum. A ``circular line node" on the largest hole FS around the Z point at the Brillouin zone (BZ) boundary was found. This result was considered as an evidence for $s\pm$ symmetry \cite{ZhangNP2012}. Alternatively, taking into account the observed correlation of the gap value with orbital character of the electronic states \cite{Evtushinsky2012}, the ``circular line node" can be explained as a location of extremely small gap due to lack of $d_{xz/yz}$ character of given FS sheet at the BZ boundary.

\begin{figure*}[t]
\begin{center}
\includegraphics[width=.8\textwidth]{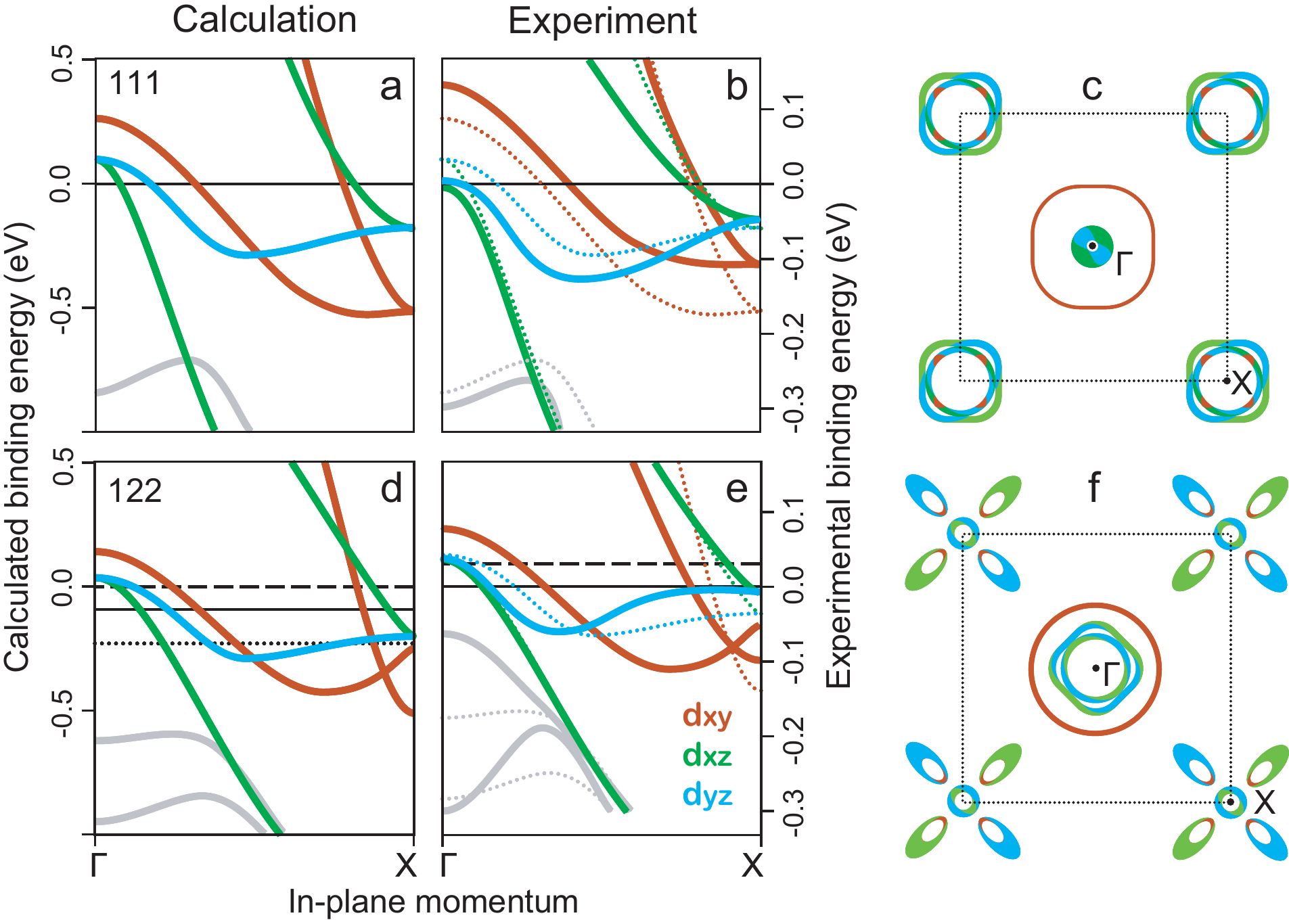}
\caption{\label{Fig_bands} Electronic band structure of LiFeAs (a-c), a representative 111 compound, and BaFe$_2$As$_2$ (BFA) / Ba$_{0.6}$K$_{0.4}$Fe$_2$As$_2$ (BKFA) (d-f), the parent/optimally doped 122 compound: the electronic bands, calculated (a, d) and derived from ARPES experiment (b, e), and the Fermi surfaces of LiFeAs (c) and BKFA (f), as seen by ARPES. The bands and FS contours are colored by the most pronounced orbital character: Fe 3$d_{xy}$, 3$d_{xz}$, and 3$d_{yz}$.}
\end{center}
\end{figure*}

\begin{figure*}[t]
\begin{center}
\includegraphics[width=.65\textwidth]{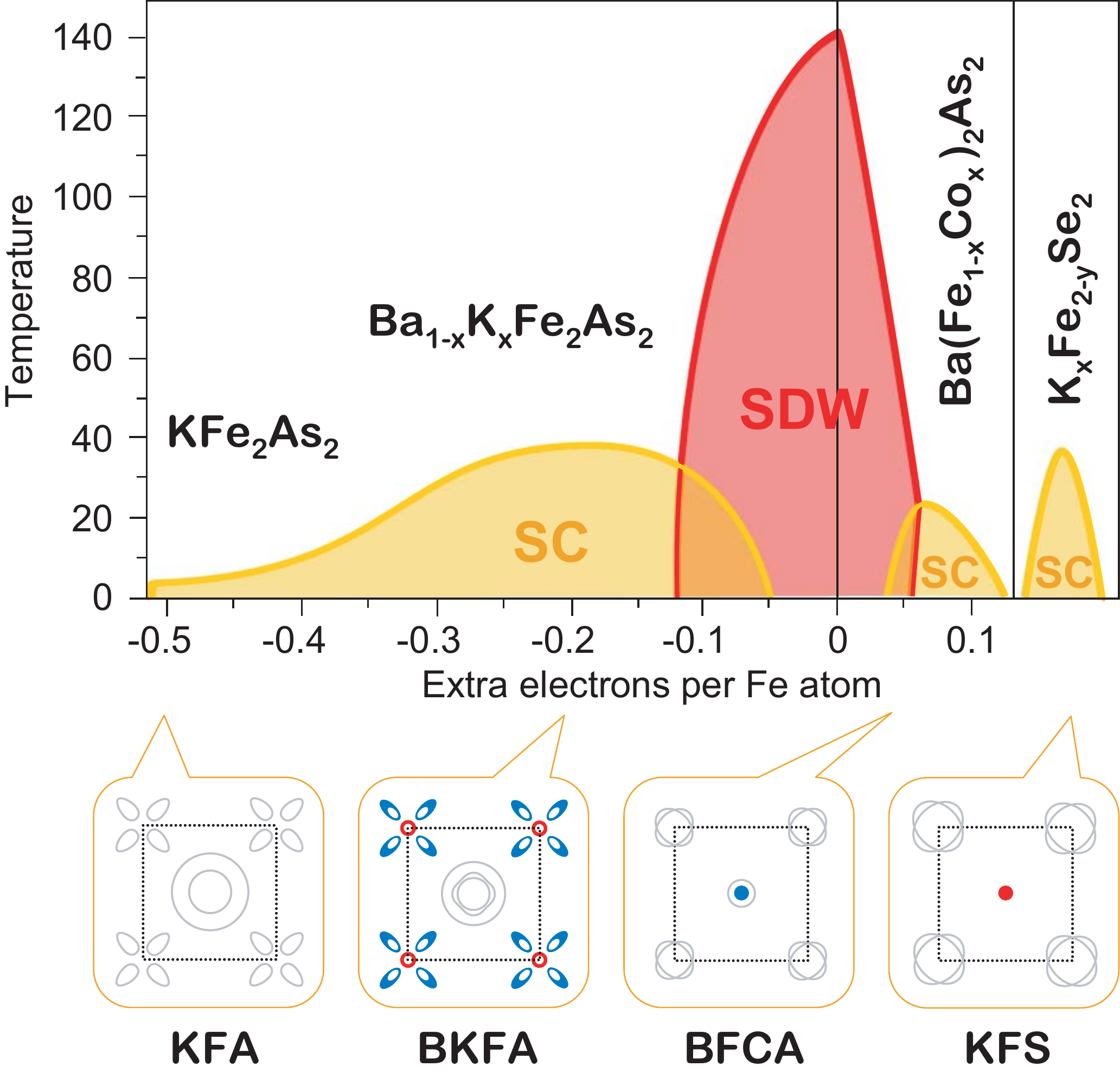}
\caption{\label{PhD}{PhD}. Phase diagram of the 122 family of ferro-pnictides complemented by the 122(Se) family as a generalized band structure driven diagram for the iron based superconductors. The insets show that the Fermi surfaces for every compound close to $T_{c\mathrm{max}}$ are in the proximity of Lifshitz topological transitions: the corresponding FS sheets are highlighted by color (blue for hole- and red for electron-like).}
\end{center}
\end{figure*}

\subsection{Electronic structure and $T_c$}

One can safely say that the visiting card of the iron based superconductors is their complex electronic band structure that usually results in five Fermi surface sheets (see Fig.\,\ref{Fig_bands}): three around the center of the Fe$_2$As$_2$ BZ and two around the corners. Band structure calculations predict rather similar electronic structure for all FeSCs (see \cite{AndersenAdP2011, Sadovskii} and references therein). ARPES experiments show that it is indeed the case: one can fit the calculated bands to the experiment if it is allowed to renormalize them about 3 times and shift slightly with respect to each other \cite{YiPRB2009, DingJoPCM2011, BorisenkoPRL2010, BorisenkoJPCS2011}. In this section, I focus first on the most ``arpesable" LiFeAs and BKFA compounds, to discuss their electronic structure in details.

\textbf{LiFeAs.} Fig.\,\ref{Fig_bands}(a) shows a fragment of the low energy electronic band structure of LiFeAs calculated using the LMTO method in the atomic sphere approximation \cite{AndersenPRB1975}. The same calculated bands but 3 times renormalized are repeated in panel (b) by the dotted lines to compare with the dispersions derived from the numerous ARPES spectra \cite{BorisenkoPRL2010, KordyukPRB2011} shown in the same panel by the thick solid lines. The experimental Fermi surface is sketched in panel (c). The five bands of interest are colored in accordance to the most pronounced orbital character: Fe 3$d_{xy}$, 3$d_{xz}$, and 3$d_{yz}$ \cite{LeePRB2008,GraserNJoP2009}. Those characters have helped us to identify uniquely the bands in the experimental spectra using differently polarized photons \cite{BorisenkoPRL2010}.

Comparing the results of the experiment and renormalized calculations, one can see that the strongest difference is observed around $\mathrm{\Gamma}$ point: the experimental $d_{xy}$ band is shifted up about 40 meV (120 meV, in terms of the bare band structure) while the $d_{xz}$/$d_{yz}$ bands are shifted about 40 (120) meV downwards. Around the corners of the BZ (X point) the changes are different, the up-shift of the $d_{xy}$ band in X point is about 60 meV while the $d_{xz}$/$d_{yz}$ bands are also shifted up slightly (about 10 meV). At the Fermi level, the largest hole-like FS sheet around $\mathrm{\Gamma}$ point, formed by $d_{xy}$ band, is essentially larger in experiment than in calculations. This is compensated by the shrunk $d_{xz}$/$d_{yz}$ FSs where the larger one has become three-dimensional, i.e. closed also in $k_z$ direction, and the smallest one has disappeared completely. The electron-like FSs have changed only slightly, alternating its character in $\mathrm{\Gamma}$X direction due to shift of the crossing of $d_{xz}$ and $d_{xy}$ bands below the Fermi level, see Fig.\,\ref{Fig_bands}(b). So, the experimental electronic band structure of LiFeAs has the following very important differences from the calculated one \cite{BorisenkoPRL2010}: (i) there is no FS nesting, see \ref{FS_Maps} (left), and (ii) the vHs, the tops of the $d_{xz}$/$d_{yz}$ bands at $\mathrm{\Gamma}$ point, stays in the vicinity of the Fermi level, i.e. the system is very close to a Lifshitz transition \cite{LifshitzZETF1960}. The latter makes the band structure of LiFeAs similar to the structure of optimally doped Ba(Fe$_{1-x}$Co$_{x})$As$_2$ (BFCA) \cite{LiuPRB2011}, as discussed below.

\textbf{BKFA.} I start from the parent stoichiometric BaFe$_2$As$_2$, a representative fragment of the calculated electronic band structure for which is shown in Fig.\,\ref{Fig_bands}(d). It is very similar to the band structure of LiFeAs with a small complication at the bottom of the $d_{xy}$ bands in X point that is a consequence of body-centered tetragonal stacking of FeAs layers instead of simple tetragonal stacking in LiFeAs.

With the highest, in 122 family, transition temperature ($T_c$ = 38~K) and the sharpest ARPES spectra, the hole doped BKFA and BNFA are the most promising and the most popular objects for trying to understand the mechanism of superconductivity in ferro-pnictides. This said, it is important to stress that the FS of the optimally doped Ba$_{0.6}$K$_{0.4}$Fe$_2$As$_2$ and Ba$_{0.6}$Na$_{0.4}$Fe$_2$As$_2$ is topologically different from the expected one: instead of two electron-like pockets around the corners of the Fe$_2$As$_2$ BZ (X and Y points) there is a propeller-like FS with the hole-like blades and a very small electron-like center \cite{ZabolotnyyN2009, ZabolotnyyPhC2009}, as shown in Fig.\,\ref{FS_Maps} (right). Curiously enough, despite the experimental reports of the propeller like FS, the ``parent'' FS is still used in a number of theoretical models and as a basis for interpretation of experimental results such as superconducting gap symmetry.

Our first interpretation of the propeller-like FS, as an evidence for an additional electronic ordering \cite{ZabolotnyyN2009}, was based on temperature dependence of the photoemission intensity around X point and on the similarity of its distribution to the parent BFA, but the interpretation based on a shift of the electronic band structure \cite{YiPRB2009} was also discussed. Now, while it seems that the electronic ordering plays a certain role in spectral weight redistribution \cite{EvtushinskyJPSJ2011}, we have much more evidence for the ``structural'' origin of the propellers: (1) The propeller-like FS, such as shown Fig.\,\ref{Fig_PropoMaps}(a), is routinely observed for every optimally doped BKFA or BNFA crystals we have studied. (2) In extremely overdoped KFA \cite{SatoPRL2009,YoshidaJPCS2011}, where the magnetic ordering is not expected at all, they naturally (according to rigid band approximation) evolve to larger hole-like propellers. (3) The same propellers in the spectrum of the overdoped ($T_c =$ 10\,K) BFCA at 90 meV below the Fermi level \cite{KordyukFPS2011}.

Fig.\,\ref{Fig_bands}(e) shows the experimental bands (solid lines), derived from a number of ARPES spectra \cite{KordyukFPS2011}, on top of the bands (thin dotted lines) calculated for parent BFA, 3 times renormalized, and shifted by 30 meV, as discussed above, to model the band structure expected for Ba$_{0.6}$K$_{0.4}$Fe$_2$As$_2$. One can see that the difference between the experimental and ``expected'' dispersions is even smaller than in case of LiFeAs and mainly appears near X point as 40 meV shifts of the $d_{xz}$/$d_{yz}$ bands and one of $d_{xy}$ bands. These small shifts, however, result in the topological Lifshitz transition of the FS and the question is how it is related to superconductivity.

Naturally, one would like to examine whether one of the peaks in the electronic density of states (DOS), related to the Lifshitz transitions, can be responsible for the enhancement of superconductivity in BKFA. Comparing the DOS calculated for the parent BFA and the model Fermi surfaces \cite{KordyukFPS2011} (see also \cite{SI}) one can see that the chemical potential, for which the FS would be the most similar to the experimental FS of BKFA, drops in the region where DOS of $d_{xz}$/$d_{yz}$ bands exhibits singularities. Strictly  speaking, at the energy of $-228$ meV DOS is not peaked but is increasing with lowering energy, hinting that a simple correlation between DOS and $T_c$, as suggested in \cite{Sadovskii}, does not work for BKFA. From this procedure one can also conclude that the extremely doped KFA should have much higher DOS than any of BKFA, that clearly contradicts to the idea of simple relation between DOS and $T_c$. On the other hand, the high $T_c$ superconductivity scenario driven by interband pairing in a multiband system in the proximity of a Lifshitz topological transition \cite{InnocentiPRB2010, InnocentiSUST2011}, looks more promising alternative for BKFA. This  said, it seems extremely challenging task for chemists to go with overdoping still further in order to reach the $d_{xz/yz}$ saddle points responsible for the largest DOS peak at $-282$ eV. Interestingly, the same can be suggested for LiFeAs, where DOS \cite{SI} shows a much higher peak of the same $d_{xz/yz}$ origin.

Going back to the Lifshitz transitions in iron based superconductors, let us overview their electronic band structures now accessible by ARPES. Recently, the correlation of the Lifshitz transition with the onset of superconductivity has been observed in BFCA \cite{LiuPRB2011,LiuNP2010}. The study has been mainly concentrated on the outer hole-like FS formed by $d_{xy}$ orbitals, nevertheless, it has been also found \cite{LiuPRB2011} that the tops of the $d_{xz}$/$d_{yz}$ bands go to the Fermi level for the samples with the optimal doping and $T_c = 24$\,K. Thus, the FS of optimally doped BFCA is similar to the FS formed by $d_{xz}$/$d_{yz}$ bands of LiFeAs, i.e. for the case when the $\mathrm{\Gamma}$-centered $d_{xz/yz}$ FS pocket is in the proximity of a Lifshitz transition. One can add another 111 compound here, NaFeAs, that also has the tops of $d_{xz}$/$d_{yz}$ bands very close to the Fermi level \cite{HePRL2010}.

One more example to support this picture comes from 245 family (see \cite{Sadovskii} and references therein). The ARPES spectra from these compounds \cite{BorisenkoXXX2012} are not very sharp yet, but one can confidently say that the bottom of the electron pocket at the center of the BZ is very close to the Fermi level, that allows us to place this family on the electron overdoped side of the generalized phase diagram, as shown in Fig.\,\ref{PhD}. At the end we note that in all known cases the bands those Lifshitz transitions do correlate with $T_c$ have predominantly Fe 3$d_{xz/yz}$ orbital character.

\section{Conclusions}

While the mechanisms of superconductivity and magnetism in FeSC remain unresolved issues, the experimental determination of electronic band structure allows us to make useful conclusions. Now we can say that the electronic structure of FeSC is either clear or can be easily clarified by experiment so that one can easily fit the calculated bands to the experiment if it is allowed to renormalize them about 3 times and shift slightly with respect to each other. So, one can suggest the following algorithm:
\begin{eqnarray}
\mathrm{experiment}&=&(\mathrm{calculation} + \mathrm{shifts}) \times \mathrm{renormalization} \nonumber\\
\mathrm{calculation}&\Rightarrow&\mathrm{orbital~character} \nonumber\\
\mathrm{shifts}&\Rightarrow&\mathrm{FS~topology} + \mathrm{nesting~conditions,} \nonumber
\end{eqnarray}
i.e.~from comparison of the experiment and calculations one can get the correct electronic structure with known orbital symmetry and estimate the self-energy (renormalization). From the former one gets the Fermi surface topology that is necessary for understanding superconductivity and FS geometry (nesting conditions) that may or may not be important for understanding the magnetism here. From renormalization one can get the information about electronic interaction.

Considering all the electronic band structures of FeSCs that can be derived from ARPES, it has been found that the Fermi surface of every optimally doped compound (the compounds with highest $T_c$) has the Van Hove singularities of the Fe 3$d_{xz/yz}$ bands in the vicinity to the Fermi level. This suggests that the proximity to an electronic topological transition, known as Lifshitz transition, for one of the multiple Fermi surfaces makes the superconductivity dome at the phase diagram. Since the parent band structure is known, one can consciously move the essential vHs to the Fermi level by charge doping, by isovalent doping, or by pressure. Based on this empirical observation, one can predict, in particular, that hole overdoping of KFe$_2$As$_2$ and LiFeAs compounds is a possible way to increase the $T_c$.

To summarize, the iron based superconductors promise interesting physics and applications. While the interplay of superconductivity and magnetism, as well as their mechanisms remain the issues of active debates and studies, one thing in FeSC puzzle is clear, namely that it is the complex multi-band electronic structure of these compounds that determines their rich and puzzling properties. What is important and fascinating is that this complexity seems to play a positive role in the struggle for understanding the FeSC physics and also for search of the materials with higher $Tc$'s. This is because the multiple electronic bands and resulting complex Fermiology offer exceptionally rich playground for establishing useful empirical correlations. This is also because this electronic structure is well understood---the band structure calculations well reproduce its complexity: all the bands and their symmetry. The role of the experiment, in this case, is just to define exact position and renormalization for each band. This piece of experimental knowledge, however, appears to be vitally important for understanding of all electronic properties of these new compounds.

\begin{acknowledgements}

I am pleased to dedicate this review to 80th anniversary of Prof. V. V. Eremenko. I acknowledge numerous discussions with members of the ARPES group at IFW Dresden: S. V. Borisenko, D. V. Evtushinsky, and V. B. Zabolotnyy, as well as with A. Bianconi, A. V. Boris, B. B\"{u}chner, A. V. Chubukov, A. M. Gabovich, G. E. Grechnev, P. J. Hirschfeld, D. S. Inosov, T. K. Kim, Yu. V. Kopaev, M. M. Korshunov, I. I. Mazin, I. V. Morozov, I. A. Nekrasov, S. G. Ovchinnikov, E. A. Pashitskii, S. M. Ryabchenko, M. V. Sadovskii, S. Thirupathaiah, M. A. Tanatar, and A. N. Yaresko. The project was supported by the DFG priority program SPP1458, Grants No. KN393/4, BO1912/2-1.
\end{acknowledgements}


\begin{thebibliography}{100}

\bibitem{KamiharaJACS2008}
{Y.~Kamihara \textit{et~al}.,
  \href{http://pubs.acs.org/doi/abs/10.1021/ja800073m}{
\newblock J. Am. Chem. Soc. {\bf 130}, 3296 (2008)}.}

\bibitem{SadovskiiPU2008}
{M.~V. Sadovskii, \href{http://stacks.iop.org/1063-7869/51/i=12/a=R02}{
\newblock Physics-Uspekhi {\bf 51}, 1201 (2008)}.}

\bibitem{IvanovskiiPU2008}
{A.~L. Ivanovskii, \href{http://stacks.iop.org/1063-7869/51/i=12/a=R03}{
\newblock Physics-Uspekhi {\bf 51}, 1229 (2008)}.}

\bibitem{IzyumovPU2008}
{Y.~A. Izyumov and E.~Z. Kurmaev,
  \href{http://stacks.iop.org/1063-7869/51/i=12/a=R04}{
\newblock Physics-Uspekhi {\bf 51}, 1261 (2008)}.}

\bibitem{IshidaJPSJ2009}
{K.~Ishida, Y.~Nakai, and H.~Hosono,
  \href{http://jpsj.ipap.jp/link?JPSJ/78/062001/}{
\newblock J. Phys. Soc. Jpn. {\bf 78}, 062001 (2009)}.}

\bibitem{MazinPhC2009}
{I.~Mazin and J.~Schmalian,
  \href{http://www.sciencedirect.com/science/article/pii/S0921453409001002}{
\newblock Physica C {\bf 469}, 614  (2009)}.}

\bibitem{JohnstonAPh2010}
{D.~C. Johnston,
  \href{http://www.tandfonline.com/doi/abs/10.1080/00018732.2010.513480}{
\newblock Advances in Physics {\bf 59}, 803 (2010)}.}

\bibitem{PaglioneNPh2010}
{J.~Paglione and R.~L. Greene, \href{http://dx.doi.org/10.1038/nphys1759}{
\newblock Nat Phys {\bf 6}, 645 (2010)}.}

\bibitem{WenARoCMP2011}
{H.-H. Wen and S.~Li,
  \href{http://www.annualreviews.org/doi/abs/10.1146/annurev-conmatphys-062910-140518}{
\newblock Annual Review of Condensed Matter Physics {\bf 2}, 121 (2011)}.}

\bibitem{StewartRMP2011}
{G.~R. Stewart, \href{http://link.aps.org/doi/10.1103/RevModPhys.83.1589}{
\newblock Rev. Mod. Phys. {\bf 83}, 1589 (2011)}.}

\bibitem{HirschfeldRPP2011}
{P.~J. Hirschfeld, M.~M. Korshunov, and I.~I. Mazin,
  \href{http://stacks.iop.org/0034-4885/74/i=12/a=124508}{
\newblock Rep. Prog. Phys. {\bf 74}, 124508 (2011)}.}

\bibitem{ChubukovAR2012}
{A.~Chubukov,
  \href{http://www.annualreviews.org/doi/abs/10.1146/annurev-conmatphys-020911-125055}{
\newblock Annual Review of Condensed Matter Physics {\bf 3}, 57 (2012)}.}

\bibitem{PuttiSST2010}
{M.~Putti \textit{et~al}.,
  \href{http://stacks.iop.org/0953-2048/23/i=3/a=034003}{
\newblock Supercond. Sci. Technol. {\bf 23}, 034003 (2010)}.}

\bibitem{MollNM2010}
{P.~J.~W. Moll \textit{et~al}., \href{http://dx.doi.org/10.1038/nmat2795}{
\newblock Nat. Mater. {\bf 9}, 628 (2010)}.}

\bibitem{GurevichRPP2011}
{A.~Gurevich, \href{http://stacks.iop.org/0034-4885/74/i=12/a=124501}{
\newblock Rep. Prog. Phys. {\bf 74}, 124501 (2011)}.}

\bibitem{PatelAPL2009}
{U.~Patel \textit{et~al}., \href{http://link.aip.org/link/?APL/94/082508/1}{
\newblock Appl. Phys. Lett. {\bf 94}, 082508 (2009)}.}

\bibitem{KordyukFPS2011}
{A.~A. {Kordyuk} \textit{et~al}., \href{http://arxiv.org/abs/1111.0288}{
\newblock arXiv:1111.0288  (2011)}.}

\bibitem{KordyukLTP2006}
{A.~A. Kordyuk and S.~V. Borisenko,
  \href{http://link.aip.org/link/?LTP/32/298/1}{
\newblock Low Temp. Phys. {\bf 32}, 298 (2006)}.}

\bibitem{NandiPRL2010}
{S.~Nandi \textit{et~al}.,
  \href{http://link.aps.org/doi/10.1103/PhysRevLett.104.057006}{
\newblock Phys. Rev. Lett. {\bf 104}, 057006 (2010)}.}

\bibitem{LuetkensNM2009}
{H.~Luetkens \textit{et~al}., \href{http://dx.doi.org/10.1038/nmat2397}{
\newblock Nat. Mater. {\bf 8}, 305 (2009)}.}

\bibitem{Katayama2010}
{N.~{Katayama} \textit{et~al}.,
  \href{http://jpsj.ipap.jp/link?JPSJ/79/113702/}{
\newblock J. Phys. Soc. Jpn. {\bf 79}, 113702 (2010)}.}

\bibitem{JiangJPCM2009}
{S.~Jiang \textit{et~al}.,
  \href{http://stacks.iop.org/0953-8984/21/i=38/a=382203}{
\newblock J. Phys.: Condens. Matter {\bf 21}, 382203 (2009)}.}

\bibitem{EschrigPRB2010}
{H.~Eschrig, A.~Lankau, and K.~Koepernik,
  \href{http://link.aps.org/doi/10.1103/PhysRevB.81.155447}{
\newblock Phys. Rev. B {\bf 81}, 155447 (2010)}.}

\bibitem{LiuPRB2010}
{C.~Liu \textit{et~al}.,
  \href{http://link.aps.org/doi/10.1103/PhysRevB.82.075135}{
\newblock Phys. Rev. B {\bf 82}, 075135 (2010)}.}

\bibitem{RotterPRL2008}
{M.~Rotter, M.~Tegel, and D.~Johrendt,
  \href{http://link.aps.org/doi/10.1103/PhysRevLett.101.107006}{
\newblock Phys. Rev. Lett. {\bf 101}, 107006 (2008)}.}

\bibitem{SefatPRL2008}
{A.~S. Sefat \textit{et~al}.,
  \href{http://link.aps.org/doi/10.1103/PhysRevLett.101.117004}{
\newblock Phys. Rev. Lett. {\bf 101}, 117004 (2008)}.}

\bibitem{NiPRB2008}
{N.~Ni \textit{et~al}.,
  \href{http://link.aps.org/doi/10.1103/PhysRevB.78.214515}{
\newblock Phys. Rev. B {\bf 78}, 214515 (2008)}.}

\bibitem{BrouetPRB2009}
{V.~Brouet \textit{et~al}.,
  \href{http://link.aps.org/doi/10.1103/PhysRevB.80.165115}{
\newblock Phys. Rev. B {\bf 80}, 165115 (2009)}.}

\bibitem{KondoPRB2010}
{T.~Kondo \textit{et~al}.,
  \href{http://link.aps.org/doi/10.1103/PhysRevB.81.060507}{
\newblock Phys. Rev. B {\bf 81}, 060507 (2010)}.}

\bibitem{RotterPRB2008}
{M.~Rotter \textit{et~al}.,
  \href{http://link.aps.org/doi/10.1103/PhysRevB.78.020503}{
\newblock Phys. Rev. B {\bf 78}, 020503 (2008)}.}

\bibitem{SatoPRL2009}
{T.~Sato \textit{et~al}.,
  \href{http://link.aps.org/doi/10.1103/PhysRevLett.103.047002}{
\newblock Phys. Rev. Lett. {\bf 103}, 047002 (2009)}.}

\bibitem{CortesGil2010}
{R.~Cortes-Gil \textit{et~al}.,
  \href{http://pubs.acs.org/doi/abs/10.1021/cm100956k}{
\newblock Chem. Mater. {\bf 22}, 4304 (2010)}.}

\bibitem{Aswartham2012}
{S.~{Aswartham} \textit{et~al}., \href{http://arxiv.org/abs/1203.0143}{
\newblock arXiv:1203.0143  (2012)}.}

\bibitem{WuJPCM2008}
{G.~Wu \textit{et~al}.,
  \href{http://stacks.iop.org/0953-8984/20/i=42/a=422201}{
\newblock J. Phys.: Condens. Matter {\bf 20}, 422201 (2008)}.}

\bibitem{ParkJPCM2008}
{T.~Park \textit{et~al}.,
  \href{http://stacks.iop.org/0953-8984/20/i=32/a=322204}{
\newblock J. Phys.: Condens. Matter {\bf 20}, 322204 (2008)}.}

\bibitem{RenPRL2009}
{Z.~Ren \textit{et~al}.,
  \href{http://link.aps.org/doi/10.1103/PhysRevLett.102.137002}{
\newblock Phys. Rev. Lett. {\bf 102}, 137002 (2009)}.}

\bibitem{RichardRPP2011}
{P.~Richard \textit{et~al}.,
  \href{http://stacks.iop.org/0034-4885/74/i=12/a=124512}{
\newblock Rep. Prog. Phys. {\bf 74}, 124512 (2011)}.}

\bibitem{LiuPRL2008}
{C.~Liu \textit{et~al}.,
  \href{http://link.aps.org/doi/10.1103/PhysRevLett.101.177005}{
\newblock Phys. Rev. Lett. {\bf 101}, 177005 (2008)}.}

\bibitem{DingEPL2008}
{H.~Ding \textit{et~al}.,
  \href{http://stacks.iop.org/0295-5075/83/i=4/a=47001}{
\newblock EPL {\bf 83}, 47001 (2008)}.}

\bibitem{ZabolotnyyN2009}
{V.~B. Zabolotnyy \textit{et~al}.,
  \href{http://www.imp.kiev.ua/~kord/papers/box/2009%20Nature%20Zabolotnyy.pdf}{
\newblock Nature {\bf 457}, 569 (2009)}.}

\bibitem{EvtushinskyPRB2009}
{D.~V. Evtushinsky \textit{et~al}.,
  \href{http://link.aps.org/doi/10.1103/PhysRevB.79.054517}{
\newblock Phys. Rev. B {\bf 79}, 054517 (2009)}.}

\bibitem{RichardPRL2009}
{P.~Richard \textit{et~al}.,
  \href{http://link.aps.org/doi/10.1103/PhysRevLett.102.047003}{
\newblock Phys. Rev. Lett. {\bf 102}, 047003 (2009)}.}

\bibitem{EvtushinskyNJP2009}
{D.~V. Evtushinsky \textit{et~al}.,
  \href{http://stacks.iop.org/1367-2630/11/i=5/a=055069}{
\newblock New J. Phys. {\bf 11}, 055069 (2009)}.}

\bibitem{EvtushinskyJPSJ2011}
{D.~V. Evtushinsky \textit{et~al}.,
  \href{http://jpsj.ipap.jp/link?JPSJ/80/023710/}{
\newblock J. Phys. Soc. Jpn. {\bf 80}, 023710 (2011)}.}

\bibitem{ThirupathaiahPRB2010}
{S.~Thirupathaiah \textit{et~al}.,
  \href{http://link.aps.org/doi/10.1103/PhysRevB.81.104512}{
\newblock Phys. Rev. B {\bf 81}, 104512 (2010)}.}

\bibitem{LiuNP2010}
{C.~Liu \textit{et~al}., \href{http://dx.doi.org/10.1038/nphys1656}{
\newblock Nat Phys {\bf 6}, 419 (2010)}.}

\bibitem{LiuPRB2011}
{C.~Liu \textit{et~al}.,
  \href{http://link.aps.org/doi/10.1103/PhysRevB.84.020509}{
\newblock Phys. Rev. B {\bf 84}, 020509 (2011)}.}

\bibitem{YiPNAS2011}
{M.~Yi \textit{et~al}.,
  \href{http://www.pnas.org/content/108/17/6878.abstract}{
\newblock PNAS {\bf 108}, 6878 (2011)}.}

\bibitem{HeumenPRL2011}
{E.~van Heumen \textit{et~al}.,
  \href{http://link.aps.org/doi/10.1103/PhysRevLett.106.027002}{
\newblock Phys. Rev. Lett. {\bf 106}, 027002 (2011)}.}

\bibitem{YiPRB2009}
{M.~Yi \textit{et~al}.,
  \href{http://link.aps.org/doi/10.1103/PhysRevB.80.024515}{
\newblock Phys. Rev. B {\bf 80}, 024515 (2009)}.}

\bibitem{FinkPRB2009}
{J.~Fink \textit{et~al}.,
  \href{http://link.aps.org/doi/10.1103/PhysRevB.79.155118}{
\newblock Phys. Rev. B {\bf 79}, 155118 (2009)}.}

\bibitem{RichardPRL2010}
{P.~Richard \textit{et~al}.,
  \href{http://link.aps.org/doi/10.1103/PhysRevLett.104.137001}{
\newblock Phys. Rev. Lett. {\bf 104}, 137001 (2010)}.}

\bibitem{KimPRB2011}
{Y.~Kim \textit{et~al}.,
  \href{http://link.aps.org/doi/10.1103/PhysRevB.83.064509}{
\newblock Phys. Rev. B {\bf 83}, 064509 (2011)}.}

\bibitem{YoshidaJPCS2011}
{T.~Yoshida \textit{et~al}.,
  \href{http://www.sciencedirect.com/science/article/pii/S0022369710003689}{
\newblock J. Phys. Chem. Solids {\bf 72}, 465  (2011)}.}

\bibitem{LiuPRL2009}
{C.~Liu \textit{et~al}.,
  \href{http://link.aps.org/doi/10.1103/PhysRevLett.102.167004}{
\newblock Phys. Rev. Lett. {\bf 102}, 167004 (2009)}.}

\bibitem{YoshidaPRL2011}
{T.~Yoshida \textit{et~al}.,
  \href{http://link.aps.org/doi/10.1103/PhysRevLett.106.117001}{
\newblock Phys. Rev. Lett. {\bf 106}, 117001 (2011)}.}

\bibitem{ZhangNP2012}
{Y.~Zhang \textit{et~al}., \href{http://dx.doi.org/10.1038/nphys2248}{
\newblock Nat. Phys. {\bf 8}, 371 (2012)}.}

\bibitem{ThirupathaiahPRB2011}
{S.~Thirupathaiah \textit{et~al}.,
  \href{http://link.aps.org/doi/10.1103/PhysRevB.84.014531}{
\newblock Phys. Rev. B {\bf 84}, 014531 (2011)}.}

\bibitem{Evtushinsky2011}
{D.~V. {Evtushinsky} \textit{et~al}., \href{http://arxiv.org/abs/1106.4584}{
\newblock arXiv:1106.4584v1  (2011)}.}

\bibitem{WangSSC2008}
{X.~Wang \textit{et~al}., \href{http://dx.doi.org/10.1016/j.ssc.2008.09.057}{
\newblock Solid State Commun. {\bf 148}, 538 (2008)}.}

\bibitem{TappPRB2008}
{J.~H. Tapp \textit{et~al}.,
  \href{http://link.aps.org/doi/10.1103/PhysRevB.78.060505}{
\newblock Phys. Rev. B {\bf 78}, 060505 (2008)}.}

\bibitem{BorisenkoPRL2010}
{S.~V. Borisenko \textit{et~al}.,
  \href{http://www.imp.kiev.ua/~kord/papers/box/2010%20PRL%20Borisenko.pdf}{
\newblock Phys. Rev. Lett. {\bf 105}, 067002 (2010)}.}

\bibitem{KordyukPRB2011}
{A.~A. Kordyuk \textit{et~al}.,
  \href{http://www.imp.kiev.ua/~kord/papers/box/2011%20PRB%20Kordyuk%20LiFeAs.pdf}{
\newblock Phys. Rev. B {\bf 83}, 134513 (2011)}.}

\bibitem{BorisenkoSym2012}
{S.~V. Borisenko \textit{et~al}., \href{http://www.mdpi.com/2073-8994/4/1/251}{
\newblock Symmetry {\bf 4}, 251 (2012)}.}

\bibitem{MorozovCGD2010}
{I.~Morozov \textit{et~al}.,
  \href{http://pubs.acs.org/doi/abs/10.1021/cg1005538}{
\newblock Cryst. Growth Des. {\bf 10}, 4428 (2010)}.}

\bibitem{ChenPRL2009}
{G.~F. Chen \textit{et~al}.,
  \href{http://link.aps.org/doi/10.1103/PhysRevLett.102.227004}{
\newblock Phys. Rev. Lett. {\bf 102}, 227004 (2009)}.}

\bibitem{LiPRB2009}
{S.~Li \textit{et~al}.,
  \href{http://link.aps.org/doi/10.1103/PhysRevB.80.020504}{
\newblock Phys. Rev. B {\bf 80}, 020504 (2009)}.}

\bibitem{TanatarPRB2012}
{M.~A. Tanatar \textit{et~al}.,
  \href{http://link.aps.org/doi/10.1103/PhysRevB.85.014510}{
\newblock Phys. Rev. B {\bf 85}, 014510 (2012)}.}

\bibitem{ParkerPRL2010}
{D.~R. Parker \textit{et~al}.,
  \href{http://link.aps.org/doi/10.1103/PhysRevLett.104.057007}{
\newblock Phys. Rev. Lett. {\bf 104}, 057007 (2010)}.}

\bibitem{HePRL2010}
{C.~He \textit{et~al}.,
  \href{http://link.aps.org/doi/10.1103/PhysRevLett.105.117002}{
\newblock Phys. Rev. Lett. {\bf 105}, 117002 (2010)}.}

\bibitem{HeJPCS2011}
{C.~He \textit{et~al}.,
  \href{http://www.sciencedirect.com/science/article/pii/S0022369710003926}{
\newblock J. Phys. Chem. Solids {\bf 72}, 479  (2011)}.}

\bibitem{SegawaJPSJ2009}
{K.~Segawa and Y.~Ando, \href{http://jpsj.ipap.jp/link?JPSJ/78/104720/}{
\newblock J. Phys. Soc. Jpn. {\bf 78}, 104720 (2009)}.}

\bibitem{SalesPRB2009}
{B.~C. Sales \textit{et~al}.,
  \href{http://link.aps.org/doi/10.1103/PhysRevB.79.094521}{
\newblock Phys. Rev. B {\bf 79}, 094521 (2009)}.}

\bibitem{HsuPNAS2008}
{F.-C. Hsu \textit{et~al}.,
  \href{http://www.pnas.org/content/105/38/14262.abstract}{
\newblock PNAS {\bf 105}, 14262 (2008)}.}

\bibitem{ImaiPRL2009}
{T.~Imai \textit{et~al}.,
  \href{http://link.aps.org/doi/10.1103/PhysRevLett.102.177005}{
\newblock Phys. Rev. Lett. {\bf 102}, 177005 (2009)}.}

\bibitem{FedorchenkoLTP2011}
{A.~V. Fedorchenko \textit{et~al}.,
  \href{http://link.aip.org/link/?LTP/37/83/1}{
\newblock Low Temp. Phys. {\bf 37}, 83 (2011)}.}

\bibitem{XiaPRL2009}
{Y.~Xia \textit{et~al}.,
  \href{http://link.aps.org/doi/10.1103/PhysRevLett.103.037002}{
\newblock Phys. Rev. Lett. {\bf 103}, 037002 (2009)}.}

\bibitem{TamaiPRL2010}
{A.~Tamai \textit{et~al}.,
  \href{http://link.aps.org/doi/10.1103/PhysRevLett.104.097002}{
\newblock Phys. Rev. Lett. {\bf 104}, 097002 (2010)}.}

\bibitem{NakayamaPRL2010}
{K.~Nakayama \textit{et~al}.,
  \href{http://link.aps.org/doi/10.1103/PhysRevLett.105.197001}{
\newblock Phys. Rev. Lett. {\bf 105}, 197001 (2010)}.}

\bibitem{MiaoPRB2012}
{H.~Miao \textit{et~al}.,
  \href{http://link.aps.org/doi/10.1103/PhysRevB.85.094506}{
\newblock Phys. Rev. B {\bf 85}, 094506 (2012)}.}

\bibitem{GuoPRB2010}
{J.~Guo \textit{et~al}.,
  \href{http://link.aps.org/doi/10.1103/PhysRevB.82.180520}{
\newblock Phys. Rev. B {\bf 82}, 180520 (2010)}.}

\bibitem{WangNC2011}
{M.~Wang \textit{et~al}., \href{http://dx.doi.org/10.1038/ncomms1573}{
\newblock Nat. Commun. {\bf 2}, 580 (2011)}.}

\bibitem{LiPRB2011}
{C.-H. Li \textit{et~al}.,
  \href{http://link.aps.org/doi/10.1103/PhysRevB.83.184521}{
\newblock Phys. Rev. B {\bf 83}, 184521 (2011)}.}

\bibitem{YanSR2012}
{Y.~J. Yan \textit{et~al}., \href{http://dx.doi.org/10.1038/srep00212}{
\newblock Sci. Rep. {\bf 2}, 212 (2012)}.}

\bibitem{QianPRL2011}
{T.~Qian \textit{et~al}.,
  \href{http://link.aps.org/doi/10.1103/PhysRevLett.106.187001}{
\newblock Phys. Rev. Lett. {\bf 106}, 187001 (2011)}.}

\bibitem{MouPRL2011}
{D.~Mou \textit{et~al}.,
  \href{http://link.aps.org/doi/10.1103/PhysRevLett.106.107001}{
\newblock Phys. Rev. Lett. {\bf 106}, 107001 (2011)}.}

\bibitem{ZhangNM2011}
{Y.~Zhang \textit{et~al}., \href{http://dx.doi.org/10.1038/nmat2981}{
\newblock Nat. Mater. {\bf 10}, 273 (2011)}.}

\bibitem{ChenPRX2011}
{F.~Chen \textit{et~al}.,
  \href{http://link.aps.org/doi/10.1103/PhysRevX.1.021020}{
\newblock Phys. Rev. X {\bf 1}, 021020 (2011)}.}

\bibitem{BorisenkoXXX2012}
{S.~V. {Borisenko} \textit{et~al}., \href{http://arxiv.org/abs/1204.1316}{
\newblock arXiv:1204.1316  (2012)}.}

\bibitem{FriemelPRB2012}
{G.~Friemel \textit{et~al}.,
  \href{http://link.aps.org/doi/10.1103/PhysRevB.85.140511}{
\newblock Phys. Rev. B {\bf 85}, 140511 (2012)}.}

\bibitem{LumsdenJPCM2010}
{M.~D. Lumsden and A.~D. Christianson,
  \href{http://stacks.iop.org/0953-8984/22/i=20/a=203203}{
\newblock Journal of Physics: Condensed Matter {\bf 22}, 203203 (2010)}.}

\bibitem{GrunerRMP1988}
{G.~Gr\"uner, \href{http://link.aps.org/doi/10.1103/RevModPhys.60.1129}{
\newblock Rev. Mod. Phys. {\bf 60}, 1129 (1988)}.}

\bibitem{DongEPL2008}
{J.~Dong \textit{et~al}.,
  \href{http://stacks.iop.org/0295-5075/83/i=2/a=27006}{
\newblock EPL {\bf 83}, 27006 (2008)}.}

\bibitem{Li2009FeTe}
{S.~Li \textit{et~al}.,
  \href{http://link.aps.org/doi/10.1103/PhysRevB.79.054503}{
\newblock Phys. Rev. B {\bf 79}, 054503 (2009)}.}

\bibitem{YildirimPRL2008}
{T.~Yildirim, \href{http://link.aps.org/doi/10.1103/PhysRevLett.101.057010}{
\newblock Phys. Rev. Lett. {\bf 101}, 057010 (2008)}.}

\bibitem{YinPRL2010}
{W.-G. Yin, C.-C. Lee, and W.~Ku,
  \href{http://link.aps.org/doi/10.1103/PhysRevLett.105.107004}{
\newblock Phys. Rev. Lett. {\bf 105}, 107004 (2010)}.}

\bibitem{SubediPRB2008}
{A.~Subedi \textit{et~al}.,
  \href{http://link.aps.org/doi/10.1103/PhysRevB.78.134514}{
\newblock Phys. Rev. B {\bf 78}, 134514 (2008)}.}

\bibitem{ProzorovPhC2009}
{R.~Prozorov \textit{et~al}.,
  \href{http://www.sciencedirect.com/science/article/pii/S0921453409001063}{
\newblock Physica C {\bf 469}, 667  (2009)}.}

\bibitem{JulienEPL2009}
{M.-H. Julien \textit{et~al}.,
  \href{http://stacks.iop.org/0295-5075/87/i=3/a=37001}{
\newblock EPL {\bf 87}, 37001 (2009)}.}

\bibitem{ParkPRL2009}
{J.~T. Park \textit{et~al}.,
  \href{http://link.aps.org/doi/10.1103/PhysRevLett.102.117006}{
\newblock Phys. Rev. Lett. {\bf 102}, 117006 (2009)}.}

\bibitem{ShermadiniPRL2011}
{Z.~Shermadini \textit{et~al}.,
  \href{http://link.aps.org/doi/10.1103/PhysRevLett.106.117602}{
\newblock Phys. Rev. Lett. {\bf 106}, 117602 (2011)}.}

\bibitem{ShenEPL2011}
{B.~Shen \textit{et~al}.,
  \href{http://stacks.iop.org/0295-5075/96/i=3/a=37010}{
\newblock EPL {\bf 96}, 37010 (2011)}.}

\bibitem{KuritaPRB2011}
{N.~Kurita \textit{et~al}.,
  \href{http://link.aps.org/doi/10.1103/PhysRevB.83.214513}{
\newblock Phys. Rev. B {\bf 83}, 214513 (2011)}.}

\bibitem{GuptaAP2006}
{L.~C. Gupta,
  \href{http://www.tandfonline.com/doi/abs/10.1080/00018730601061130}{
\newblock Advances in Physics {\bf 55}, 691 (2006)}.}

\bibitem{KopaevZETF1970}
{Y.~V. Kopaev,
\newblock Zh. Eksp. Teor. Fiz. {\bf 58}, 1012 (1970)
\newblock [Sov. Phys. JETP. 31544 (1970)].}

\bibitem{MattisPRL1970}
{D.~C. Mattis and W.~D. Langer,
  \href{http://link.aps.org/doi/10.1103/PhysRevLett.25.376}{
\newblock Phys. Rev. Lett. {\bf 25}, 376 (1970)}.}

\bibitem{RusinovJETP1973}
{A.~I. Rusinov, D.~C. Kat, and K.~Y. V.,
\newblock Zh. Eksp. Teor. Fiz. {\bf 65}, 1984 (1973)
\newblock [Sov. Phys. JETP 38, 991 (1974)].}

\bibitem{BalseiroPRB1979}
{C.~A. Balseiro and L.~M. Falicov,
  \href{http://link.aps.org/doi/10.1103/PhysRevB.20.4457}{
\newblock Phys. Rev. B {\bf 20}, 4457 (1979)}.}

\bibitem{KopaevPLA1987}
{Y.~Kopaev and A.~Rusinov,
  \href{http://www.sciencedirect.com/science/article/pii/0375960187905330}{
\newblock Phys. Lett. A {\bf 121}, 300  (1987)}.}

\bibitem{GabovichLTP2000}
{A.~M. Gabovich and A.~I. Voitenko,
  \href{http://link.aip.org/link/?LTP/26/305/1}{
\newblock Low Temp. Phys. {\bf 26}, 305 (2000)}.}

\bibitem{GabovichACMP2010}
{A.~M. Gabovich \textit{et~al}., \href{http://dx.doi.org/10.1155/2010/681070}{
\newblock Advances in Condensed Matter Physics {\bf 2010} (2010)}.}

\bibitem{FangPRB2005}
{L.~Fang \textit{et~al}.,
  \href{http://link.aps.org/doi/10.1103/PhysRevB.72.014534}{
\newblock Phys. Rev. B {\bf 72}, 014534 (2005)}.}

\bibitem{MorosanNP2006}
{E.~Morosan \textit{et~al}., \href{http://dx.doi.org/10.1038/nphys360}{
\newblock Nat. Phys. {\bf 2}, 544 (2006)}.}

\bibitem{Kato1988}
{M.~Kato and K.~Machida,
  \href{http://link.aps.org/doi/10.1103/PhysRevB.37.1510}{
\newblock Phys. Rev. B {\bf 37}, 1510 (1988)}.}

\bibitem{DahmNP2009}
{T.~Dahm \textit{et~al}., \href{http://dx.doi.org/10.1038/nphys1180}{
\newblock Nat. Phys. {\bf 5}, 217 (2009)}.}

\bibitem{KordyukEPJ2010}
{A.~Kordyuk \textit{et~al}.,
  \href{http://dx.doi.org/10.1140/epjst/e2010-01303-3}{
\newblock Eur. Phys. J. Special Topics {\bf 188}, 153 (2010)}.}

\bibitem{NingPRL2010}
{F.~L. Ning \textit{et~al}.,
  \href{http://link.aps.org/doi/10.1103/PhysRevLett.104.037001}{
\newblock Phys. Rev. Lett. {\bf 104}, 037001 (2010)}.}

\bibitem{MatanPRB2010}
{K.~Matan \textit{et~al}.,
  \href{http://link.aps.org/doi/10.1103/PhysRevB.82.054515}{
\newblock Phys. Rev. B {\bf 82}, 054515 (2010)}.}

\bibitem{WakimotoJPSJ2010}
{S.~Wakimoto \textit{et~al}., \href{http://jpsj.ipap.jp/link?JPSJ/79/074715/}{
\newblock Journal of the Physical Society of Japan {\bf 79}, 074715 (2010)}.}

\bibitem{IikuboJPSJ2009}
{S.~Iikubo \textit{et~al}., \href{http://jpsj.ipap.jp/link?JPSJ/78/103704/}{
\newblock Journal of the Physical Society of Japan {\bf 78}, 103704 (2009)}.}

\bibitem{LesterPRB2010}
{C.~Lester \textit{et~al}.,
  \href{http://link.aps.org/doi/10.1103/PhysRevB.81.064505}{
\newblock Phys. Rev. B {\bf 81}, 064505 (2010)}.}

\bibitem{LiPRB2010}
{H.-F. Li \textit{et~al}.,
  \href{http://link.aps.org/doi/10.1103/PhysRevB.82.140503}{
\newblock Phys. Rev. B {\bf 82}, 140503 (2010)}.}

\bibitem{ArgyriouPRB2010}
{D.~N. Argyriou \textit{et~al}.,
  \href{http://link.aps.org/doi/10.1103/PhysRevB.81.220503}{
\newblock Phys. Rev. B {\bf 81}, 220503 (2010)}.}

\bibitem{DialloPRB2010}
{S.~O. Diallo \textit{et~al}.,
  \href{http://link.aps.org/doi/10.1103/PhysRevB.81.214407}{
\newblock Phys. Rev. B {\bf 81}, 214407 (2010)}.}

\bibitem{VignolleNP2007}
{B.~Vignolle \textit{et~al}., \href{http://dx.doi.org/10.1038/nphys546}{
\newblock Nat. Phys. {\bf 3}, 163 (2007)}.}

\bibitem{MonthouxPRL1994}
{P.~Monthoux and D.~J. Scalapino,
  \href{http://link.aps.org/doi/10.1103/PhysRevLett.72.1874}{
\newblock Phys. Rev. Lett. {\bf 72}, 1874 (1994)}.}

\bibitem{KorshunovPRB2008}
{M.~M. Korshunov and I.~Eremin,
  \href{http://link.aps.org/doi/10.1103/PhysRevB.78.140509}{
\newblock Phys. Rev. B {\bf 78}, 140509 (2008)}.}

\bibitem{MaierPRB2008}
{T.~A. Maier and D.~J. Scalapino,
  \href{http://link.aps.org/doi/10.1103/PhysRevB.78.020514}{
\newblock Phys. Rev. B {\bf 78}, 020514 (2008)}.}

\bibitem{ChristiansonN2008}
{A.~D. Christianson \textit{et~al}.,
  \href{http://dx.doi.org/10.1038/nature07625}{
\newblock Nature {\bf 456}, 930 (2008)}.}

\bibitem{LumsdenPRL2009}
{M.~D. Lumsden \textit{et~al}.,
  \href{http://link.aps.org/doi/10.1103/PhysRevLett.102.107005}{
\newblock Phys. Rev. Lett. {\bf 102}, 107005 (2009)}.}

\bibitem{InosovNP2010}
{D.~S. Inosov \textit{et~al}., \href{http://dx.doi.org/10.1038/nphys1483}{
\newblock Nat. Phys. {\bf 6}, 178 (2010)}.}

\bibitem{LeePRB2010}
{S.-H. Lee \textit{et~al}.,
  \href{http://link.aps.org/doi/10.1103/PhysRevB.81.220502}{
\newblock Phys. Rev. B {\bf 81}, 220502 (2010)}.}

\bibitem{ParkPRL2011}
{J.~T. Park \textit{et~al}.,
  \href{http://link.aps.org/doi/10.1103/PhysRevLett.107.177005}{
\newblock Phys. Rev. Lett. {\bf 107}, 177005 (2011)}.}

\bibitem{InosovPRB2007}
{D.~S. Inosov \textit{et~al}.,
  \href{http://link.aps.org/doi/10.1103/PhysRevB.75.172505}{
\newblock Phys. Rev. B {\bf 75}, 172505 (2007)}.}

\bibitem{OnariPRB2010}
{S.~Onari, H.~Kontani, and M.~Sato,
  \href{http://link.aps.org/doi/10.1103/PhysRevB.81.060504}{
\newblock Phys. Rev. B {\bf 81}, 060504 (2010)}.}

\bibitem{TimuskRPP1999}
{T.~Timusk and B.~Statt, \href{http://stacks.iop.org/0034-4885/62/i=1/a=002}{
\newblock Reports on Progress in Physics {\bf 62}, 61 (1999)}.}

\bibitem{BorisenkoPRL2008}
{S.~V. Borisenko \textit{et~al}.,
  \href{http://link.aps.org/doi/10.1103/PhysRevLett.100.196402}{
\newblock Phys. Rev. Lett. {\bf 100}, 196402 (2008)}.}

\bibitem{KordyukPRB2009}
{A.~A. Kordyuk \textit{et~al}.,
  \href{http://link.aps.org/doi/10.1103/PhysRevB.79.020504}{
\newblock Phys. Rev. B {\bf 79}, 020504 (2009)}.}

\bibitem{Huefner2008}
{S.~Huefner \textit{et~al}.,
  \href{http://stacks.iop.org/0034-4885/71/i=6/a=062501}{
\newblock Rep. Prog. Phys. {\bf 71}, 062501 (2008)}.}

\bibitem{TanatarPRB2010}
{M.~A. Tanatar \textit{et~al}.,
  \href{http://link.aps.org/doi/10.1103/PhysRevB.82.134528}{
\newblock Phys. Rev. B {\bf 82}, 134528 (2010)}.}

\bibitem{SolovevLTP2009}
{A.~L. Solov'ev \textit{et~al}., \href{http://link.aip.org/link/?LTP/35/826/1}{
\newblock Low Temp. Phys. {\bf 35}, 826 (2009)}.}

\bibitem{SatoJPSJ2008}
{T.~Sato \textit{et~al}., \href{http://jpsj.ipap.jp/link?JPSJ/77/063708/}{
\newblock J. Phys. Soc. Jpn. {\bf 77}, 063708 (2008)}.}

\bibitem{HaiYun2008}
{H.-Y. Liu \textit{et~al}.,
  \href{http://159.226.36.45/Jwk_cpl/EN/abstract/article_42665.shtml}{
\newblock Chin. Phys. Lett. {\bf 25}, 3761 (2008)}.}

\bibitem{XuNC2011}
{Y.-M. Xu \textit{et~al}., \href{http://dx.doi.org/10.1038/ncomms1394}{
\newblock Nat. Commun. {\bf 2}, 392 (2011)}.}

\bibitem{ShimojimaSci2011}
{T.~Shimojima \textit{et~al}.,
  \href{http://www.sciencemag.org/content/332/6029/564.abstract}{
\newblock Science {\bf 332}, 564 (2011)}.}

\bibitem{Evtushinsky2012}
{D.~V. {Evtushinsky} \textit{et~al}., \href{http://arxiv.org/abs/1204.2432v1}{
\newblock arXiv:1204.2432v1  (2012)}.}

\bibitem{YinPhC2009}
{Y.~Yin \textit{et~al}.,
  \href{http://www.sciencedirect.com/science/article/pii/S0921453409000914}{
\newblock Physica C {\bf 469}, 535  (2009)}.}

\bibitem{MasseeEPL2010}
{F.~Massee \textit{et~al}.,
  \href{http://stacks.iop.org/0295-5075/92/i=5/a=57012}{
\newblock EPL {\bf 92}, 57012 (2010)}.}

\bibitem{SefatRPP2011}
{A.~S. Sefat, \href{http://stacks.iop.org/0034-4885/74/i=12/a=124502}{
\newblock Rep. Prog. Phys. {\bf 74}, 124502 (2011)}.}

\bibitem{HardyEPL2010}
{F.~Hardy \textit{et~al}.,
  \href{http://stacks.iop.org/0295-5075/91/i=4/a=47008}{
\newblock EPL {\bf 91}, 47008 (2010)}.}

\bibitem{PopovichPRL2010}
{P.~Popovich \textit{et~al}.,
  \href{http://link.aps.org/doi/10.1103/PhysRevLett.105.027003}{
\newblock Phys. Rev. Lett. {\bf 105}, 027003 (2010)}.}

\bibitem{OkabePRB2010}
{H.~Okabe \textit{et~al}.,
  \href{http://link.aps.org/doi/10.1103/PhysRevB.81.205119}{
\newblock Phys. Rev. B {\bf 81}, 205119 (2010)}.}

\bibitem{MazinPRL2008}
{I.~I. Mazin \textit{et~al}.,
  \href{http://link.aps.org/doi/10.1103/PhysRevLett.101.057003}{
\newblock Phys. Rev. Lett. {\bf 101}, 057003 (2008)}.}

\bibitem{OnariPRL2009}
{S.~Onari and H.~Kontani,
  \href{http://link.aps.org/doi/10.1103/PhysRevLett.103.177001}{
\newblock Phys. Rev. Lett. {\bf 103}, 177001 (2009)}.}

\bibitem{NakaiPRB2010}
{Y.~Nakai \textit{et~al}.,
  \href{http://link.aps.org/doi/10.1103/PhysRevB.81.020503}{
\newblock Phys. Rev. B {\bf 81}, 020503 (2010)}.}

\bibitem{YamashitaPRB2011}
{M.~Yamashita \textit{et~al}.,
  \href{http://link.aps.org/doi/10.1103/PhysRevB.84.060507}{
\newblock Phys. Rev. B {\bf 84}, 060507 (2011)}.}

\bibitem{VorontsovPRB2010}
{A.~B. Vorontsov, M.~G. Vavilov, and A.~V. Chubukov,
  \href{http://link.aps.org/doi/10.1103/PhysRevB.81.174538}{
\newblock Phys. Rev. B {\bf 81}, 174538 (2010)}.}

\bibitem{KontaniPRL2010}
{H.~Kontani and S.~Onari,
  \href{http://link.aps.org/doi/10.1103/PhysRevLett.104.157001}{
\newblock Phys. Rev. Lett. {\bf 104}, 157001 (2010)}.}

\bibitem{YanagiPRB2010}
{Y.~Yanagi, Y.~Yamakawa, and Y.~\ifmmode~\bar{O}\else \={O}\fi{}no,
  \href{http://link.aps.org/doi/10.1103/PhysRevB.81.054518}{
\newblock Phys. Rev. B {\bf 81}, 054518 (2010)}.}

\bibitem{Khodas2012}
{M.~Khodas and A.~V. Chubukov, \href{http://arxiv.org/abs/1202.5563}{
\newblock Phys. Rev. Lett.  (2012)}
\newblock (arXiv:1202.5563).}

\bibitem{HankePRL2012}
{T.~H\"anke \textit{et~al}.,
  \href{http://link.aps.org/doi/10.1103/PhysRevLett.108.127001}{
\newblock Phys. Rev. Lett. {\bf 108}, 127001 (2012)}.}

\bibitem{AndersenAdP2011}
{O.~K. Andersen and L.~Boeri, \href{http://dx.doi.org/10.1002/andp.201000149}{
\newblock Annalen der Physik {\bf 523}, 8 (2011)}.}

\bibitem{Sadovskii}
{M.~V. Sadovskii, E.~Z. Kuchinskii, and I.~A. Nekrasov,
  \href{http://arxiv.org/abs/1106.3707}{
\newblock arXiv:1106.3707v1  (2011)}.}

\bibitem{DingJoPCM2011}
{H.~Ding \textit{et~al}.,
  \href{http://stacks.iop.org/0953-8984/23/i=13/a=135701}{
\newblock J. Phys.: Condens. Matter {\bf 23}, 135701 (2011)}.}

\bibitem{BorisenkoJPCS2011}
{S.~V. Borisenko \textit{et~al}.,
  \href{http://www.sciencedirect.com/science/article/pii/S0022369710003677}{
\newblock J. Phys. Chem. Solids {\bf 72}, 562  (2011)}.}

\bibitem{AndersenPRB1975}
{O.~K. Andersen, \href{http://link.aps.org/doi/10.1103/PhysRevB.12.3060}{
\newblock Phys. Rev. B {\bf 12}, 3060 (1975)}.}

\bibitem{LeePRB2008}
{P.~A. Lee and X.-G. Wen,
  \href{http://link.aps.org/doi/10.1103/PhysRevB.78.144517}{
\newblock Phys. Rev. B {\bf 78}, 144517 (2008)}.}

\bibitem{GraserNJoP2009}
{S.~Graser \textit{et~al}.,
  \href{http://stacks.iop.org/1367-2630/11/i=2/a=025016}{
\newblock New Journal of Physics {\bf 11}, 025016 (2009)}.}

\bibitem{LifshitzZETF1960}
{I.~M. Lifshitz,
\newblock Zh. Eksp. Teor. Fiz. {\bf 38}, 1569 (1960)
\newblock [Sov. Phys. JETP \textbf{11}, 1130 (1960)].}

\bibitem{ZabolotnyyPhC2009}
{V.~B. Zabolotnyy \textit{et~al}.,
  \href{http://www.sciencedirect.com/science/article/pii/S0921453409000793}{
\newblock Physica C: Superconductivity {\bf 469}, 448  (2009)}.}

\bibitem{InnocentiPRB2010}
{D.~Innocenti \textit{et~al}.,
  \href{http://link.aps.org/doi/10.1103/PhysRevB.82.184528}{
\newblock Phys. Rev. B {\bf 82}, 184528 (2010)}.}

\bibitem{InnocentiSUST2011}
{D.~Innocenti \textit{et~al}.,
  \href{http://stacks.iop.org/0953-2048/24/i=1/a=015012}{
\newblock Supercond. Sci. Technol. {\bf 24}, 015012 (2011)}.}

\bibitem{SI}
{For details, see \href{http://www.imp.kiev.ua/~kord/papers/FPS11}{
\newblock www.imp.kiev.ua/\textasciitilde kord/papers/FPS11}.}

\end{thebibliography}

\end{document}